\documentclass[12pt,preprint]{aastex}

\slugcomment{submitted to Astrophysical Journal}
\shorttitle{LSB rotation curves}
\shortauthors{de Blok}

\begin{document}
\title{Halo Mass Profiles and Low Surface Brightness galaxies rotation curves}
\author{W.J.G.~de~Blok}
\affil{Research School of Astronomy and Astrophysics}
\affil{Australian National University, Mount Stromlo Observatory}
\affil{Cotter Road, Weston Creek, ACT 2611, Australia}
\email{edeblok@mso.anu.edu.au}

\begin{abstract}
A recent study has claimed that the rotation curve shapes and mass
densities of Low Surface Brightness (LSB) galaxies are largely
consistent with $\Lambda$CDM predictions, in contrast to a large body
of observational work. I demonstrate that the method used to derive
this conclusion is incapable of distinguishing the characteristic
steep CDM mass-density distribution from the core-dominated
mass-density distributions found observationally: even core-dominated
pseudo-isothermal haloes would be inferred to be consistent with CDM.
This method can therefore make no definitive statements on the
(dis)agreement between the data and CDM simulations.  After
introducing an additional criterion that does take the slope of the
mass-distribution into account I find that only about a quarter of the
LSB galaxies investigated are possibly consistent with CDM. However,
for most of these the fit parameters are so weakly constrained that
this is not a strong conclusion.  Only 3 out of 52 galaxies have
tightly constrained solutions consistent with $\Lambda$CDM. Two of
these galaxies are likely dominated by stars, leaving only one
possible dark matter dominated, CDM-consistent candidate, forming a
mere 2 per cent of the total sample. These conclusions are based on
comparison of data and simulations at identical radii and fits to the
entire rotation curves. LSB galaxies that are consistent with CDM
simulations, if they exist, seem to be rare indeed.
\end{abstract}

\keywords{galaxies: dwarf - galaxies: fundamental parameters -
  galaxies: kinematics and dynamics - dark matter}

\section{Introduction}

In recent years the inner structure of dark matter haloes has been the
topic of some discussion. Observations suggest that the dark matter
distribution in disk galaxies has a roughly constant density core,
with a typical size of order a few kpc. A recent analysis suggests a
shallow power-law mass density distribution $\rho(r) \sim r^{\alpha}$
with $\alpha \simeq -0.2 \pm 0.2$ \citep{blok2003}.  On the other
hand, numerical simulations based on the ($\Lambda$)Cold Dark Matter
(CDM) paradigm suggest a very steep inner mass density distribution, a
so-called ``cusp'' \citep{NFW96,NFW97} [NFW].  The most recent
simulations suggest slopes with $-1.5 \la \alpha \la -1$
\citep{fuku,klypin,ghigna,moore98,moore99,jing}. There is strong
consensus that this slope cannot be as shallow as the observations
suggest.

The rotation curves of Low Surface Brightness (LSB) galaxies are
considered particularly clean tests of the CDM paradigm at galaxy
scales. LSB galaxies are dominated by dark matter \citep{blok97}, and
their dynamics should thus give a more or less direct map of the dark
matter distribution.  LSB galaxy rotation curves, together with the
missing dwarf or substructure problem (e.g.\ \citealt{moore99}),
currently form the most serious challenges for the CDM model.  There
has been much debate on this topic in recent years.  This is however
not the place to review this debate, and the reader is referred to the
summaries by e.g.\ \citet{binney2004,blok2003,swaters2003} and
\citet{blok2004} and references therein.

On the theoretical side the ``universal mass density profile'' (with
its characteristic $\alpha = -1$ cusp) as proposed by
\citet{NFW96,NFW97}, has long been a corner-stone of CDM (but see
\citealt{moore98,moore99}).  Recent simulations presented in
\citet{NFW04} and \citet{hayashi} [hereafter H04] show that simulated
CDM haloes may actually contain a range of slopes in their inner
regions (though the slopes remain steep).  The shallowest inner slope
measured is $\alpha \sim -1$, with an average over the simulated
haloes of $\alpha \sim -1.2$ (at the innermost reliably resolved
radius for galaxy mass haloes).  With this scatter in mind, H04 argue
that simple fitting functions (such as ``the'' NFW halo) do not
capture the full variety and diversity in shapes and slopes, and
suggest that this may partly explain the discrepancy between
observations and simulations.  They claim that the inner slope is
difficult to constrain, both observationally as well as theoretically
and point out that rotation curve constraints are strongest where
numerical simulations are least reliable.  This, as well as the
potential for systematic effects in the observations, has, at least
according to H04, led to an ``unwarranted emphasis'' on the value of
the inner slope, at the cost of evaluation of the data and simulations
over their full radial extent.

H04 present a novel method to fit rotation curves derived from their
simulations as well as LSB galaxy rotation curves from the
literature. Their method is able to describe the variety present in
data and simulations and from the distributions of respective fitting
parameters H04 conclude that the observed curves \emph{are}
consistent with the CDM paradigm; a conclusion that contradicts a
large body of observational work.

For their analysis, H04 make use of the rotation curves presented in
\citet{mcgaugh2001}, \citet{dBB02} and \citet{swaters2003}, and derive
circular velocity curves for haloes from their simulations.  To
quantify both sets of observed and simulated rotation curves they use
a three-parameter fitting formula  (see
e.g.\ \citealt{courteau}):

\begin{equation}
V(r) =
{{V_0}\over{\left(1+\left({{r_t}\over{r}}\right)^{\gamma}\right)^{1/\gamma}}}.
\end{equation}

Here $V_0$ is the asymptotic velocity of the flat part of the rotation
curve, $r_t$ is a scale radius (the transition radius between the
rising and flat part of the rotation curve) and
the parameter $\gamma$ describes the abruptness of the turn-over
between the rising and flat parts of the rotation curve. A higher
value of $\gamma$ results in an abrupt transition, a low value in a
more gradual turn-over. Note that this function was designed to fit
observed rotation curves: it specifically assumes a solid-body rise in
the inner parts and a flat rotation curve in the outer parts.

H04 fit the two sets of rotation curves with this function,
constraining the fit parameters slightly by insisting that
\begin{enumerate} 
\item
$0<\gamma\leq 5$,
\item $r_t>0$,
\item $V_0 \leq 2V_{\rm max}$, 
\end{enumerate} with
$V_{\rm max}$ the maximum observed velocity in the rotation curve.
Fits with $\gamma > 5$ [rejected by condition (1)] correspond with
very abrupt transitions between the rising and flat parts of the
curve. Condition (3) prevents some of the solid-body rotation curves
where the asymptotic velocity is not well-constrained, from producing
fits with extremely large $V_0$.  Conditions (1)--(3) thus help
removing unrealistic values for the fit parameters.

\begin{figure}
\plotone{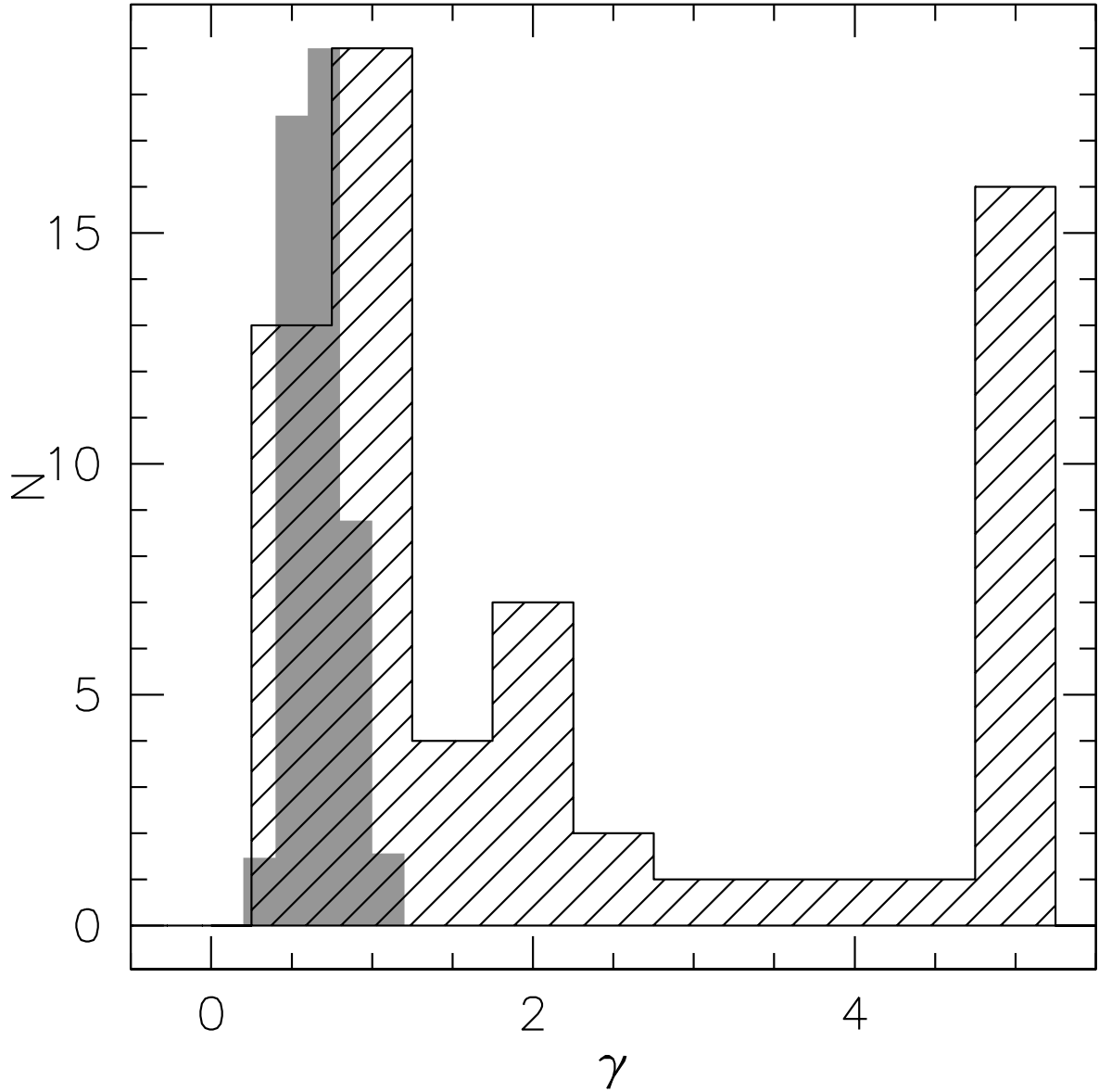}
\caption{Histogram of $\gamma$-values derived in H04. The hatched
  histogram shows the distribution derived for the LSB rotation
  curves; the grey histogram shows the distribution for the
  simulations presented in H04. The latter histogram has been
  arbitrarily scaled so that the maxima match.
\label{fig:hayashidata}} \end{figure}

Figure \ref{fig:hayashidata} (after H04, their Fig.~9) shows the
results presented in H04.  The LSB rotation curves show a broad
distribution in $\gamma$ with two pronounced peaks: one around $\gamma
\sim 1$ and a narrower one at $\gamma \sim 5$. Some 70 per cent of the
fits have $\gamma \la 2$. The $\gamma$ distribution of the simulated
CDM haloes is markedly different.  It is narrow and centered around
$\gamma \sim 0.6$ with a dispersion of $\sim 0.4$.

At first sight the two distributions seem quite incompatible, and one
might conclude that the LSB rotation curves are not consistent with
the CDM simulations.  H04 argue however that for most of the
LSB galaxy rotation curves the reduced $\chi^2$ distribution is quite
broad and shallow, so that by constraining the fit parameters slightly
more, a fit can be found that is consistent with the simulations, at
the cost of only a small increase in $\chi^2_{\rm red}$.  They define these
CDM-compatible constraints as:
\begin{enumerate}
\renewcommand{\labelenumi}{\roman{enumi}.}
\item $0<\gamma\leq 1$
\item $r_t>0$
\item $V_0 \leq 2V_{\rm max}$
\item $\left|\log
\Delta_{V/2} - \log
\Delta_{V/2,{\rm CDM}}\right| \leq 0.7$.
\end{enumerate}

Here $\Delta_{V/2}$ is the average density within the radius where the
velocity reaches half its maximum value (Alam et al 2002).  See H04
for a justification of these additional constraints, here it is
sufficient to note that for the LSB galaxies $\Delta_{V/2}$ is derived
from the fit of Eq.~(1) to the rotation curves, whereas
$\Delta_{V/2,{\rm CDM}}$ is the predicted value for a galaxy with
$V_{\rm max} = V_0$ in the $\Lambda$CDM model presented in
\citet{bullock01} and \citet{wechsler02} (see also Fig.~2 in
\citealt{alam02} and Fig.~11 in H04). The constraint on $\Delta_{V/2}$
ensures that only fits are considered with average densities close to
those predicted in a standard $\Lambda$CDM universe.

H04 find that for the majority ($\sim$ 70 per cent) of the galaxies
fits can be found for which both the best-fit $\chi^2_{\rm red} < 1.5$
[using constraints (1)-(3)] and the constrained $\chi^2_{\rm red,CDM}
< 1.5$ [using constraints (i)-(iii)]. This group of galaxies (Group A
using the terminology in H04) is thus considered to be consistent with
CDM. For the remaining galaxies no CDM-compatible fit could be found
(Group B in H04), or the curves were too irregular to be well fitted
by Eq.~(1) (Group C in H04).

The main claim in H04 is thus that most of the LSB galaxies can still
be fitted reasonably well with these extra constraints and without a
large increase in $\chi^2_{\rm red}$.  They conclude, based on the
agreements of the constrained, CDM-compatible $\gamma$ and
$\Delta_{V/2}$ distributions with those of the simulations, that
``this sample of LSB rotation curves is not manifestly inconsistent
with the predictions of $\Lambda$CDM cosmological models.''  This
conclusion thus contradicts most observational work on LSB rotation
curves which argues the opposite.

In this paper I discuss these conclusions and investigate the method
with which they are derived.  I find that the method presented in H04
cannot distinguish between shallow and steep mass-density slopes, and
illustrate this by showing that with this method even core-dominated,
pseudo-isothermal halos would be inferred to be consistent with CDM.
I introduce an additional criterion that does take the inner
mass-density slope into account, and find that only a quarter of the
LSB galaxies investigated are possibly consistent with CDM. However,
most of these galaxies' fit parameters are so weakly constrained that
this is not a very strong conclusion.  Only 3 out of the 51 galaxies
investigated have tightly constrained solutions that are not
significantly inconsistent with CDM.  However, two of them are high
surface brightness galaxies that are likely dominated by their stellar
population. The net result is that in the total sample there is only
one galaxy that is likely to be dark matter dominated and not
significantly inconsistent with CDM. LSB galaxies that are consistent
with CDM seem to be rare indeed.

In Sect.~2 I derive an analytical expression for the slope of a
rotation curve, based on Eq.~(1). Sect.~3 compares the slopes in both
data and simulations at identical physical radii. In Sect.~4 I show
that the constraints imposed by H04 are not sufficient to prove or
disprove agreement with CDM. In Sect.~5 a new constraint is introduced
that does take the slope into account. The results are summarised in
Sect.~6.

\section{Slopes}

\begin{figure}
\plotone{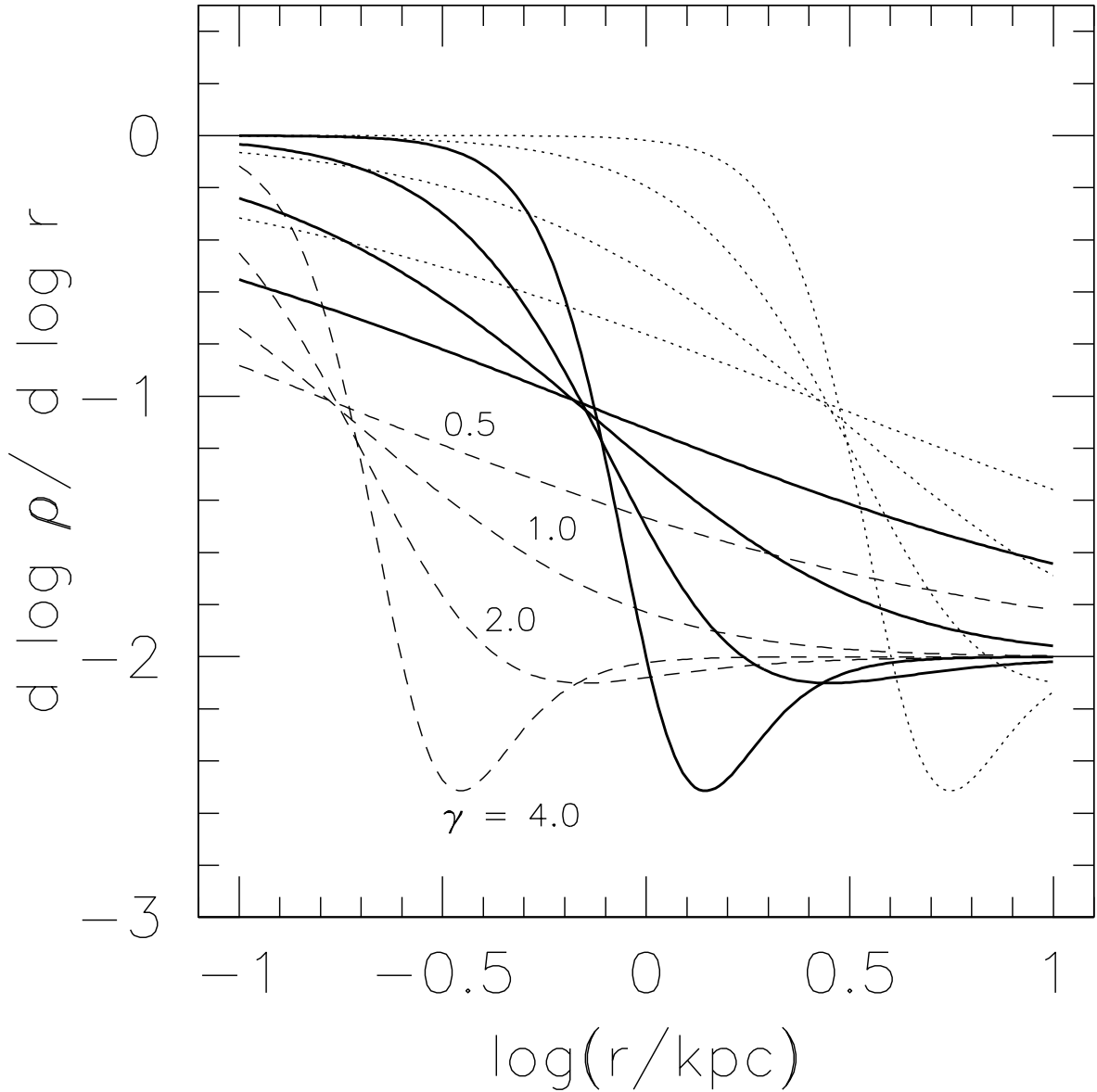}
\caption{Logarithmic slopes (Eq.~4) of the mass 
density distributions (Eq.~3) corresponding to the rotation curves
given in Eq.~(1). From left to right, the dashed curves have $r_t =
0.25$ kpc, the solid curves $r_t = 1$ kpc, and the dotted curves $r_t
= 4$ kpc, respectively. For each value of $r_t$, curves with $\gamma =
(0.5, 1, 2, 4)$ are plotted as indicated. Note that the value of the
slope at any radius depends on both $\gamma$ and $r_t$.
\label{fig:courslope}} \end{figure}

Equation~(1) is a flexible, analytical function that can describe a
wide range of rotation curves (see \citealt{courteau} for more
examples).  One can, analogous to H04, use the fitting function to
investigate the properties of the underlying data.  Assuming a
spherical halo dominated by dark matter we can use the inversion used
in e.g.\ \citet{blok2001b} to derive the corresponding mass-density
distribution:

\begin{equation}
4 \pi G \rho(r) = 2 {{V}\over{r}} \left({{dV}\over{dr}}\right)+ \left({{V}\over{r}}\right)^2,
\end{equation}
in combination with Eq.~(1) yields

\begin{equation}
4 \pi G \rho(r) = \left({{V_0}\over{r}}\right)^2 
\left[{{ 1 + 3\left( {{r_t}\over{r}} \right)^{\gamma} \hfil}\over{{\left( 1 +
\left( {{r_t}\over{r}} \right)^{\gamma} \right)^{(2+\gamma)/\gamma}}}}\right].
\end{equation}
The logarithmic slope of the mass density
distribution can now be derived from Eq.~(3):

\begin{equation} 
\alpha(r) \equiv {{d \log \rho}\over{d \log r}} = -{{2+\gamma}\over{1+\left(
{{r_t}\over{r}} \right)^{\gamma}}}  + {{\gamma}\over{1+3\left(
{{r_t}\over{r}}\right)^{\gamma}}}.
\end{equation}
The logarithmic, or power-law slope $\alpha(r)$ thus \emph{depends on
  both $\gamma$ and $r_t$}.  Fig.~\ref{fig:courslope} shows
$\alpha(r)$ for a number of values of $\gamma$ and $r_t$, where the
values have been chosen to cover the range shown by the LSB rotation
curve fits from H04. Changing $\gamma$ has the effect of changing the
shape of the slope-radius relations; changing $r_t$ shifts the
relations horizontally without affecting the shape for any particular
$\gamma$.  Most of the plotted slopes are remarkably \emph{shallow} at
small radii.  An inner asymptotic value of $\alpha = 0$ is
predominant.  The CDM slope $\alpha \sim -1$ can only be reproduced
with Eqs.~(1) and (4) for small values of $r_t$ and/or extremely small
values of $\gamma$. Similarly, at large radii the slope converges to
$\alpha = -2$.  These are not naturally occurring values for
$\Lambda$CDM mass profiles, with asymptotic slopes $\alpha \la -1$ in
the inner parts and $\alpha=-3$ in the outer parts.  Equation (1) is
really built to mimic the same striking features of rotation curves
that led to the introduction of the pseudo-isothermal (ISO) halo,
namely a solid-body rise in the inner parts, and a constant (``flat'')
rotation velocity in the outer parts. One needs to make very specific
choices of fitting parameters to make this function resemble an NFW
profile.

\section{Comparing Observations and Simulations}

The H04 simulations predict steep slopes, even at the smallest
reliably resolved radii, defined by the so-called convergence radius
$r_{\rm conv}$ \citep{power03}. For the H04 dwarf galaxy models
$r_{\rm conv}= 0.3h^{-1}$ kpc or $\sim 0.4$ kpc for $h=0.7$. The slope
at $r_{\rm conv}$ in the simulations varies between $\alpha \sim -1$
and $\sim -1.3$.

Equation~(4) can be used to derive the slopes implied by the fits to
the observed LSB rotation curves.  For this one needs to choose a
radius at which the slopes will be evaluated. In the rest of this
paper two choices will be used. Firstly, the radius of the innermost
measured point of the rotation curve as given in \citet{blok2001b} and
\citet{dBB02} will be considered\footnote{\citet{swaters2003} do not
  provide values for $r_{\rm in}$. Here a value of $2''$ is used
  (converted using the appropriate distance), which is the typical
  spacing between data points for the curves presented there.}.  One
might argue that at $r= r_{\rm in}$ a different radius is probed for
each rotation curve, so, secondly, a constant radius $r = 0.4$ kpc $
\simeq r_{\rm conv}$ kpc for all curves will be used as well.  One
should not expect dramatically different results from both choices
though: the average value of $r_{\rm in}$ is $\sim 0.45$ kpc, very
much comparable with $r_{\rm conv}$.  It should be kept in mind that
even though we evaluate the slope at these particular choices for the
radius, the $\gamma$ and $r_t$ parameters (and therefore the slope)
are constrained by the \emph{entirety} of the curve.

\begin{figure}
\plotone{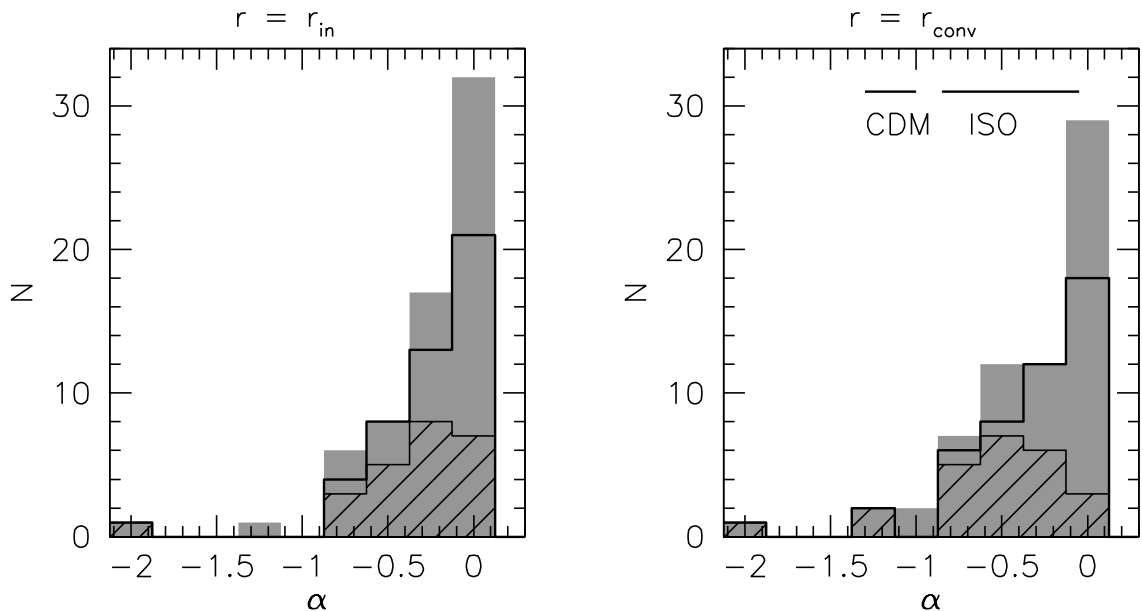}
\caption{Histogram of the inner slopes derived from the best
  $(\gamma,r_t)$ fits for $r=r_{\rm in}$ (left panel) and $r=r_{\rm
    conv}$ (right panel). The grey histogram shows the distribution
  for the entire sample, the open histogram that for Group A alone,
  and the hatched histogram for the subset of Group A for which the
  best fit is apparently already consistent with CDM. In the right
  hand panel the line labeled ``CDM'' marks the range of slopes found
  at this radius in the H04 simulations. The line marked ``ISO''
  indicated the range of slopes expected for ISO halos at this radius
  (cf.\ Fig.~\ref{fig:slopeonly}).
\label{fig:hayslope3}} \end{figure}

Figure \ref{fig:hayslope3} shows the distribution of inferred mass
density slopes derived using the LSB rotation curves and the best-fit
$\gamma$ and $r_t$ parameters listed in Table 2 of H04. The slopes for
both $r=r_{\rm in}$ and $r=r_{\rm conv}=0.4$ kpc are shown. It is
clear that the distributions are heavily biased towards shallow slopes
and very different from the values found in the simulations at these
radii $(\alpha \la -1)$.  The small values of $\chi^2_{\rm red, best}$
derived in H04 indicate that Eq.~(1) fits the data well, and the
problem is thus not with the quality of the fits.  In
Fig.~\ref{fig:hayslope3} is also indicated the distribution of slopes
for the galaxies that H04 claim are consistent with CDM (their Group A
with $\chi^2_{\rm red,best} <1.5$ and $\chi^2_{\rm red,CDM} <
1.5$). This distribution is not markedly different from that of the
entire sample.  From Fig.~\ref{fig:hayslope3} one could thus conclude
that the majority of slopes are inconsistent with the CDM prediction.
How can this be reconciled with the H04 conclusions?

As discussed above, H04 add an extra constraint to the
$\gamma$-parameter, and argue that the resulting agreement with the
range of $\gamma$-values shown by the simulations implies consistency
with CDM.  This point will be explored in more detail below, but one
can already see here that this does not solve the problem of the
discrepancy between observed and simulated slopes.  Consider the
distribution of slopes of the subset of Group A for which $\chi^2_{\rm
red,best} = \chi^2_{\rm red,CDM}$ (and both $< 1.5$), i.e. galaxies
for which the best fit is (apparently) already consistent with
CDM. The distribution of slopes of this sub-group is also plotted in
Fig.~\ref{fig:hayslope3}. The selection criteria for this sub-group
favour galaxies with steeper slopes, and it should thus come as no
surprise that the dominant peak at $\alpha \sim 0$ has
disappeared. Nevertheless it is remarkable that the distributions are
still dominated by slopes $\alpha > -1$. The distribution using $r =
r_{\rm in}$ peaks at $\alpha \sim -0.25$. The distribution using $r =
0.4$ kpc peaks at $\alpha \sim -0.5$. As we are probing the slope at
$r=0.4$ kpc and not in the centre, we should not expect all ISO haloes
to have a flat slope at this radius.  The distribution of slopes found
is consistent with that expected at $r = 0.4$ kpc for an ensemble of
pseudo-isothermal (ISO) halos with core-radii between $\sim 0.5$ and
$\sim 5$ kpc\footnote{Note that this range is consistent with the one
derived from direct ISO rotation curve fits. For example, the 5th and
95th percentile values of the core-radius distribution for the minimum
disk ISO models listed in
\citet{dBB02} and \citet{mcgaugh2001} are 0.6 and 5.0, respectively.}. The decrease in
mass-density at this radius is thus much less steep than expected for
a CDM halo. The steepest slope expected for a realistic ISO halo
(i.e.\ consistent with the smallest value $R_C \sim 0.5$ kpc measured
in real LSB galaxies; see results in
\citealt{dBB02, mcgaugh2001,swaters2003}) is $\alpha \sim -0.8$, with
the large majority of ISO haloes having slopes less steep than that.
The slopes measured in the H04 simulations at $r = 0.4$ kpc vary
between $\sim -1$ and $\sim -1.3$, firmly inconsistent with the
observed distributions.

The reason for the discrepancy can best be appreciated by realizing
that according to Eq.~(4) the slope is a function of both $\gamma$ and
$r_t$. This is shown graphically in Fig.~\ref{fig:gammartfits} which
shows iso-slope contours in the $(\gamma, r_t)$ plane, as evaluated at
$r=0.4$ kpc. It is immediately obvious that $\alpha$ is not a unique
function of $\gamma$, but depends equally strongly on $r_t$. Most
combinations of $\gamma$ and $r_t$ with $\gamma \ga 1.5$ and $r_t \ga
1.5$ yield slopes shallower than $\alpha = -0.2$. Only a very small
part of parameter space results in steep, CDM-compatible slopes
(essentially only the area with $r_t < 0.6$ and/or $\gamma <
0.2$). Constraining $\gamma <1$ does not actually constrain the slope
to steep, CDM-like values, but still allows values up to $\alpha \sim
-0.2$.

Over-plotted in Fig.~\ref{fig:gammartfits} are the best-fit $\gamma$
and $r_t$ values of the LSB rotation curves as determined by H04. A
distinction is made between the entire sample (Groups A, B and C), the
``CDM-compatible'' Group A, and the sub-group discussed above where
the best fit is apparently already CDM-compatible. The distribution of
the points is not very different from group to group, though the
$\chi^2_{\rm red, best} = \chi^2_{\rm red, CDM}$ subgroup tends to
have slightly steeper slopes, as already shown in
Fig.~\ref{fig:hayslope3}.  Regardless of how the total sample is
divided up, the very large majority of the galaxies have slopes
inconsistent with the values $\alpha \la -1$ derived in the H04
simulations. Note again that the slopes in simulations and
observations are compared at \emph{identical radii}, and that the
$\gamma$ and $r_t$ values were derived using the \emph{entirety of the
rotation curves}.  The extra constraints introduced by H04 are thus
insufficient conditions for agreement with CDM, as constraining to
$\gamma <1$ does not automatically imply a mass-density slope
consistent with CDM. Unless one is prepared to allow for shallow
slopes in CDM at these radii (making it inconsistent with its own
simulations) an extra constraint on the slope is needed to ensure
agreement with CDM.

\section{Reductio ad absurdum with ISO haloes}

Adding this extra constraint to the four introduced by H04 is only
necessary if these prior constraints are insufficient to make a unique
distinction between CDM and non-CDM models.  If, for example, rotation
curves known to be inconsistent with CDM are not rejected as such,
the prior constraints are clearly insufficient to address CDM-related
questions.

One can test the strength of these constraints  by ``observing'' a
sample of simulated ISO haloes.  This will show that the H04
constraints are not sufficient as they imply that the majority of
these ISO haloes, all with easily detectable shallow slopes, are
consistent with CDM. ISO haloes form of course the one model known to
be certainly inconsistent with CDM (if they were not, there would 
be no cusp/core controversy).  The modeling procedure as described in
\citet{blok2003} [hereafter dB03] is used.  The reader is referred to
that paper for an extensive description of these models. The aim there
was to estimate the impact of systematic effects on the observations.
For that purpose many different rotation curves were simulated and
``observed'' after the addition of observational uncertainties due to
e.g.\ resolution effects and random motions. dB03 used the same
datasets as the basis of their models as H04 use for their analysis,
and the dB03 procedure can be directly applied here to test the
fitting method and make comparisons with the H04 results.

The ISO halo parameters are chosen in such a way that the resulting
curves resemble minimum disk ISO fits to the observed rotation
curves. Fig.~10 from \citet{blok2001a} shows that mass models of LSB
galaxies have asymptotic velocities $1.8 < \log (V_{\infty}/[{\rm km\
s}^{-1}]) < 2.2$ (between $\sim 60$ and $\sim 160$ km s$^{-1}$).
Central dark matter core densities vary between $-2 < \log
(\rho_0/[M_{\odot}\,{\rm pc}^{-3}]) < 0$ (between $10^{-2}$ and $1$
$M_{\odot}$ pc$^{-3}]$). The core radius $R_C$ follows from these two
parameters, and has a range $-0.57 < \log (R_C/[{\rm kpc}]) < 0.83$
(between $0.27$ and $6.8$ kpc).  For the models a random $V_{\infty}$
and $\rho_0$ are chosen from the respective logarithmic intervals.
Note that these intervals encompass the entire observed parameter
range. No account is taken of any correlations that might exist
between these parameters, nor of any other uncertainties that might be
present in the determination of halo parameters. This is reflected in
the range of core-radii, which is somewhat larger than that found from
observations (typically $0.5 \la R_C \la 5$ kpc). The choice of
parameter ranges is therefore liberal, but sufficient for our
purposes: this is not an attempt to model individual galaxies but a
\emph{reductio ad absurdum} proof of principle.


For each model halo an inclination,  resolution and sampling interval
are chosen randomly from the observed distributions described in dB03.
Error-bars are assigned to each sampled point, again modeled on the
observed distribution of uncertainties, with a minimum value of 4 km
s$^{-1}$.  The curve is then corrected for inclination with a
corresponding increase in the size of the error-bars.  To each data
point a random velocity component between $-10$ and $10$ km s$^{-1}$
is added. This is slightly simpler than the method used in dB03, but
results in similarly sized random motions. Lastly, as real rotation
curves do not all extend out to the same radius, for each galaxy a
random outer radius between 3 and 20 kpc is defined, out to which the
galaxy is ``observed''.

Figure \ref{fig:panel} shows a small selection of the artificial ISO
curves.  Over-plotted is the Eq.~(1) fit using the best-fit
constraints (1)-(3), as given in Sec.~1. The reduced $\chi^2$ and the
fit parameters are also given in the figure.
Figure~\ref{fig:modelgamma} shows the distribution of the best-fit
$\gamma$ parameters for 100 different realisations of the model.  Some
$\sim 80$ per cent of the ISO galaxies have $\gamma \la 2$. This
number should be compared with H04 where it was found that $(70 \pm
5)$ per cent of the LSB rotation curves had $\gamma \la 2$. The
$\gamma$-distribution of LSB galaxies is at least in this respect
consistent with that of ISO haloes.  Fig.~\ref{fig:modelslopes} shows
a histogram of the slopes at $r=0.4$ kpc, both for the input ISO
models and the ``observed'' output models, the latter derived using
the best-fit $\gamma$ and $r_t$ parameters.  Both the input and output
slopes show a clear peak at $\alpha = 0$ with a tail towards steeper
slopes where the data are probing the edge of the dark matter
core\footnote{Note again that as we are probing the slope at $r=0.4$
kpc and not in the center, we do not expect to find flat slopes for
all ISO haloes. The steepest ISO slopes expected are nevertheless
significantly less steep than the predicted CDM values. Cf.\ the
remarks in Sect.~3.}.  The output distribution has a less pronounced
peak at $\alpha = 0$, and a longer tail towards steeper slopes, but as
the discussion in \citealt{blok2001b} shows, this is fully consistent
with the effects of the limited resolution used to ``observe'' the
input distribution. The range in slopes found is slightly larger than
that exhibited by real LSB galaxies shown in
Fig.~\ref{fig:hayslope3}. This is entirely due to the larger range in
core-radii used.  Fits of Eq.~(1) to a set of model ISO rotation
curves ``observed'' under the same conditions as the samples of LSB
galaxies under investigation here are thus able to retain and retrieve
the presence of shallow inner slopes.

In order for the H04 criteria to make a valid discrimination between
CDM and non-CDM models they thus need to be able to reject the
majority of the ISO models presented here.  Applying these criteria to
the ISO models and again fitting using Eq.~(1), one finds that
reducing the upper limit of $\gamma = 5$ to $\gamma = 1$ has only a
modest effect on $\chi^2_{\rm red}$ and almost the entire ISO sample
can still be fitted reasonably well with Eq.~(1) and constraints
(i)-(iii), as can be seen from the representative examples in
Fig.~\ref{fig:panel}.  CDM constraint (iv) turns out to be the most
restrictive one. It rejects about $\sim 40$ per cent of the simulated
ISO haloes with $\left|\log \Delta_{V/2} - \log \Delta_{V/2,{\rm
    CDM}}\right| > 0.7$ (note that H04 show that a similar fraction of
the LSB galaxies have incompatible densities as well).  The bottom row
of Fig.~\ref{fig:panel} shows a few of the rejected models. These do
not look any different from the CDM-compatible ISO curves: condition
(iv) merely removes some fraction of the ISO haloes that do not have a
CDM-compatible density, without any particular selection against
specific values of $\gamma$, $r_t$, $V_0$ or other relevant
parameters.  The most surprising outcome though is that of this sample
of random ISO haloes almost 60 per cent match the H04 CDM criteria.  A
comparison of the distribution of slopes in this CDM-compatible sample
of haloes with the total distribution shows that there has been no
selection against shallow slopes (Fig.~\ref{fig:slopescomp3}). Of the
total sample of 100 observed ISO models, $84$ per cent has an observed
slope $\alpha > -0.8$.  Once condition (iv) is applied, 57 haloes
remain, of which 82 per cent have best-fit slopes $\alpha >
-0.8$. Even when the slopes are derived using the constrained ``CDM
compatible'' fit parameters, one still finds that 70 per cent has a
shallow slope $\alpha > -0.8$.  As the range in core-radii used for
the models is slightly larger than found observationally, the
fractions mentioned here should be considered lower limits.
Regardless of how the slope is defined, shallow slopes are still
profusely present (and detectable!)  in the models.

A comparison with the input ISO halo parameters shows that the rejected
haloes are not abnormal in any way.  In Fig.~\ref{fig:rho0}
$\Delta_{V/2}$ is plotted against the input ISO halo central density
$\rho_0$. There is a clear relation, as might be expected: the average
density of a halo is related to its central density. Galaxies are
plotted with different symbols depending on whether they meet the
$\Delta_{V/2}$ constraint or not.  It is clear that the halo models
that do not meet the CDM criteria are only guilty of having average
densities that are too high or too low. They are not degenerate, nor
do they suffer from e.g.\ larger error bars.  A more finely-tuned
choice of halo parameters could bring the rejection rate down
significantly, but as the input ranges of $\rho_0$ and $V_\infty$ are
based on observed distributions, it is more likely that real galaxies
have a wider range of densities than predicted by CDM, and condition
(iv) merely selects from a much broader spectrum of galaxy properties
those galaxies that happen to lie in the CDM-compatible range.

Figure \ref{fig:chi2} compares the values of $\chi^2_{\rm red}$ for
the best-fit parameters and the CDM-compatible parameters. Haloes that
were rejected by the $\Delta_{V/2}$ criterion are indicated in the
right-hand side of the plot.  Following H04 one can divide the haloes
in three categories: \emph{(Group A):} rotation curves that are well
fit by Eq.\ (1) and for which a good fit with apparently CDM-compatible
parameters can also be found, i.e. $\chi^2_{\rm red, best} < 1.5$ and
$\chi^2_{\rm red, CDM} < 1.5$.  \emph{(Group B):} rotation curves that
are well fit by Eq. (1) but for which no good fit with CDM-compatible
parameters can be found, usually because of a discrepant value of
$\Delta_{V/2}$; these curves have $\chi^2_{\rm best} < 1.5$ and
$\chi^2_{\rm red, CDM} \ge 1.5$ or undefined (i.e., with
$\Delta_{V/2}$ out of range).  \emph{(Group C):} rotation curves that
are poorly fit by Eq.~(1), i.e. $\chi^2_{\rm red, best} \ge 1.5$ and
$\chi^2_{\rm red, CDM} \ge 1.5$.  The latter category only contains a
few galaxies, as there are no large-scale asymmetries in the model
rotation curves, and most rise smoothly and monotonically with radius.
Note the numerical agreement between the observational $\chi^2_{\rm
  red}$ values presented in H04 and those of the ISO models, showing
that the additional uncertainties introduced in those models are a
good representation of the true observational uncertainties.

Models in Group A are, according to H04, ``consistent with
$\Lambda$CDM in terms of their inferred central densities and the
shape of their rotation curves.''  One now has to add several caveats
to this statement. Firstly, even with $\gamma<1$ and $\chi^2_{\rm
  red,CDM}< 1.5$ a steep slope is not guaranteed. The constraints used
to define Group A are not sufficient to filter out the shallow slopes
that are still present and detectable in the ISO models, even after
the addition of observational uncertainties. Secondly, as the inner
slopes are well-defined and not necessarily steep, the shapes of the
rotation curves are also not necessarily consistent with
$\Lambda$CDM. Constraints (i)-(iii) thus seem to have had hardly any
effect in improving the agreement with CDM. The only restrictive
constraint is condition (iv), the $\Delta_{V/2}$ criterion, but even
this seems to have only selected models with a certain range in
average density from amongst a much larger range of models. This is
illustrated in Fig.~\ref{fig:bullock} where a comparison is made
between average $\Delta_{V/2}$ densities of the ISO models from Group
A and B with those expected in a $\Lambda$CDM universe.  Galaxies in
Groups A and B are indicated separately. In the H04 interpretation the
ISO halos from Group A are consistent with CDM in terms of shape and
average density, while group B is consistent in terms of shape but not
density, even though Groups A and B both consist of pure ISO haloes,
with easily detectable shallow slopes.  The CDM constraints (i)-(iv)
fail to take this into account and are thus insufficient to
distinguish between CDM and non-CDM models.  They cannot exclude the
possibility that LSB galaxies have shallow inner mass-density slopes,
even at $r=r_{\rm conv}$.  Additional constraints are needed to make a
unique identification of galaxies compatible with CDM.

\section{Additional Constraints}

Now that it is established that constraints (i)-(iv) are not
sufficient to uniquely identify CDM-compatible galaxies, the next step
is to identify additional conditions that can.
Figures~\ref{fig:courslope} and \ref{fig:gammartfits} show that the
inner slope and therefore the shape of the rotation curve do not
depend on $\gamma$ alone, but that there is an equally strong
dependence on $r_t$.  These Figures also explain why H04 came to the
conclusion that their four constraints were sufficient: implicit in
their analysis is the assumption that the shape of the rotation curve
depends on $\gamma$ alone.  While it is true that for $r_t \ga 0.6$
kpc a smaller $\gamma$ implies a steeper slope, this does not ensure a
steep slope in an absolute sense. For $\gamma <1$, depending on the
value of $r_t$, slopes as shallow as $\alpha =-0.2$ are still
possible.  Furthermore, it is shown above that the H04 method does not
take these shallow slopes into account, and thus allows models with
easily detectable and prominent shallow slopes purpose-built in to
pass the CDM test with flying colours.

\begin{figure}
\plotone{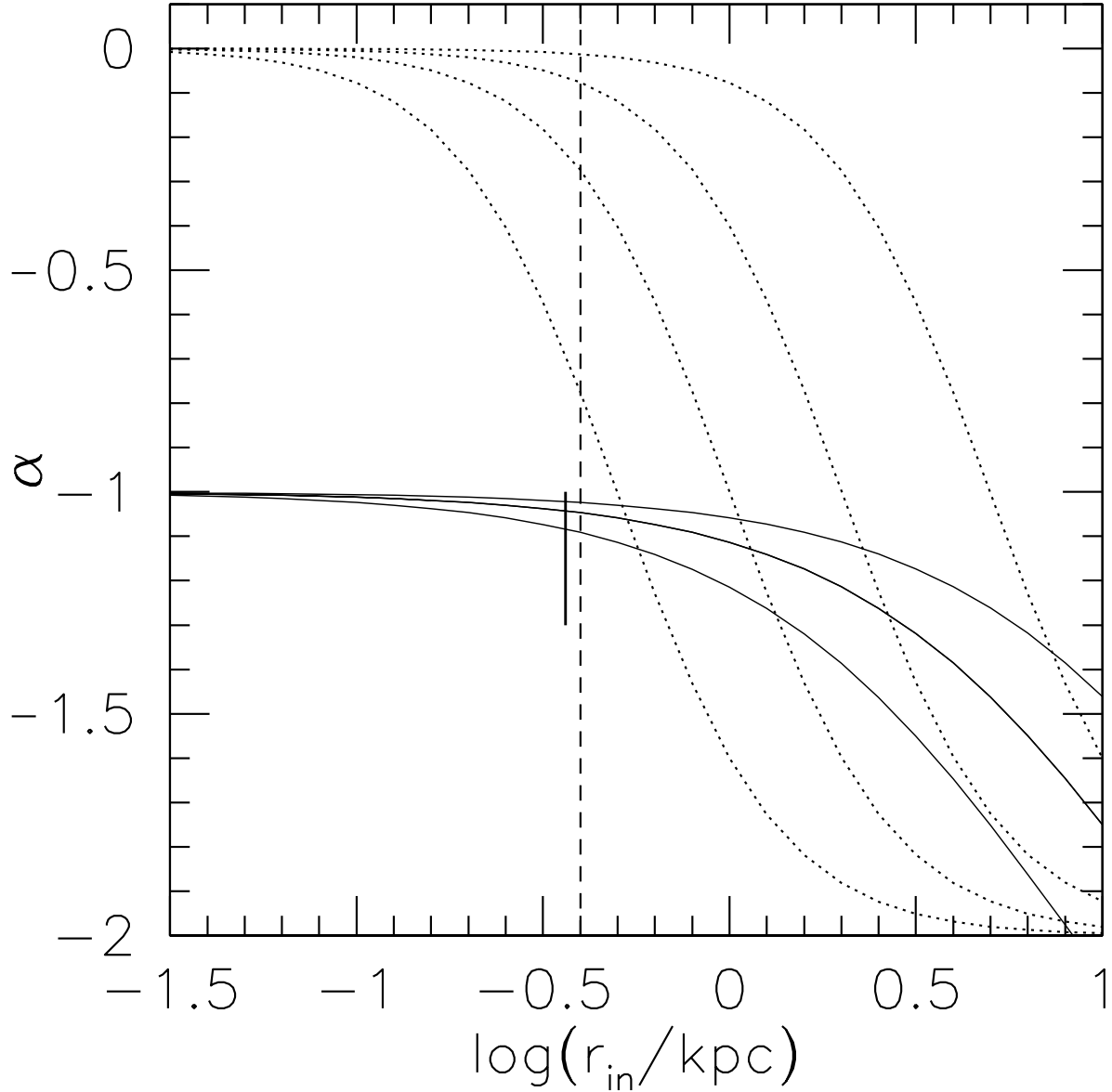}
\caption{The expected mass-density slope as a function of radius for a
  set of ISO models (dotted curves) and NFW models (full curves).
  From left to right, the ISO models have core radii $R_C = 0.5, 1, 2,
  5$ kpc. The NFW models from left to right have values $c/V_{200} =
  8/50, 8/100, 8/200$. The vertical dashed line marks the convergence
  radius $r_{\rm conv} = 0.4$ kpc. The thick vertical full line marks
  the range of slopes found in the H04 models at this radius.
\label{fig:slopeonly}} \end{figure}

\begin{figure}
\plotone{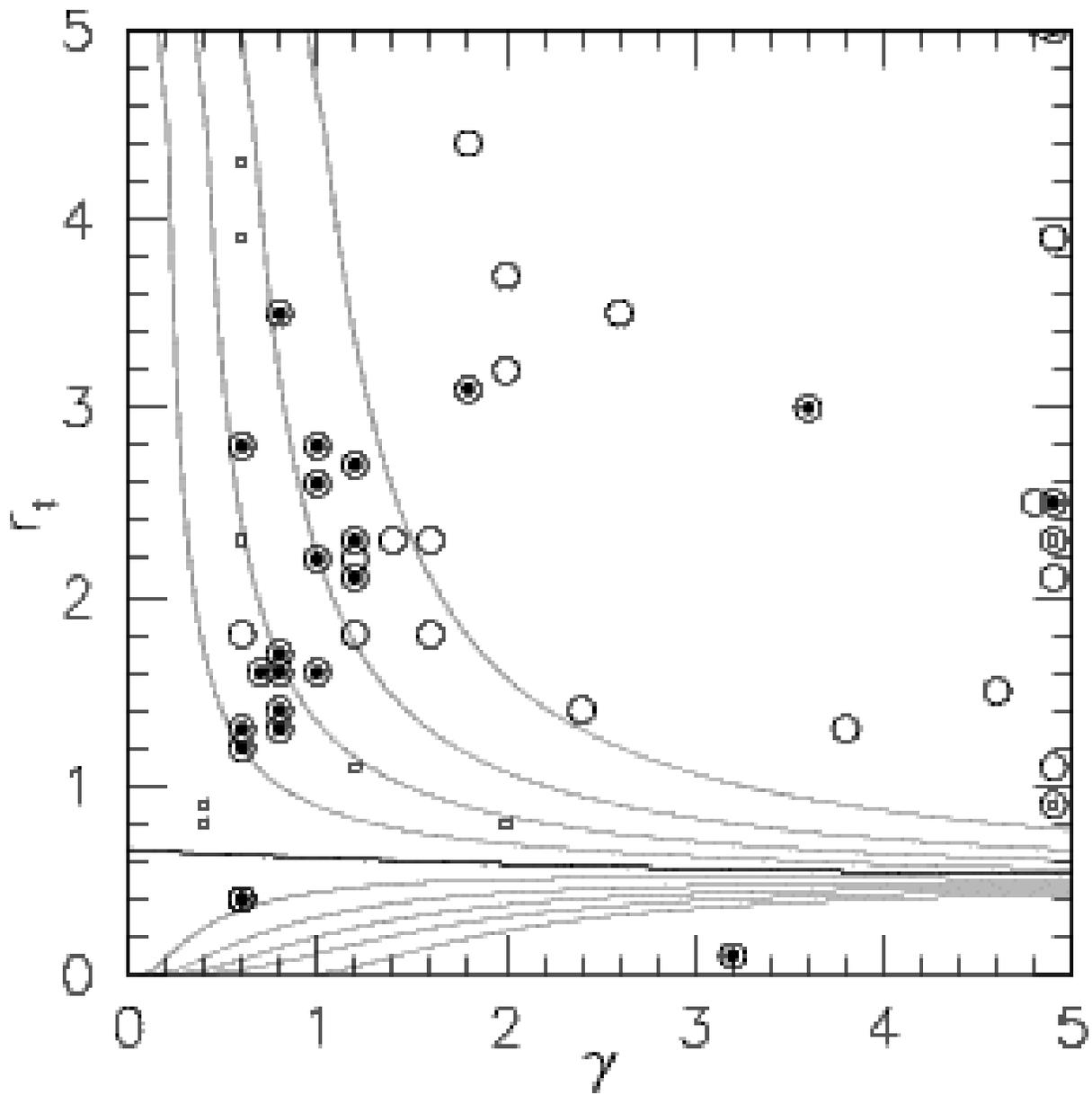}
\caption{Distribution of slopes $\alpha$ in the $(\gamma,r_t)$ plane.
  Iso-slope contours run from $\alpha = -0.2$ (top contour) to $\alpha
  = -2.0$ (bottom contour) in steps of $\Delta \alpha = -0.2$. The
  black, almost horizontal contour indicates a slope $\alpha = -1$.
  Over-plotted are the best-fit $(\gamma,r_t)$-combinations for the LSB
  galaxies in Group A (large open circles) and Groups B and C (small
  squares). Galaxies from Group A with $\chi^2_{\rm red,best} =
  \chi^2_{\rm red,CDM}$ are additionally marked with a small filled
  circle.
\label{fig:gammartfits}} \end{figure}

\begin{figure}
\plotone{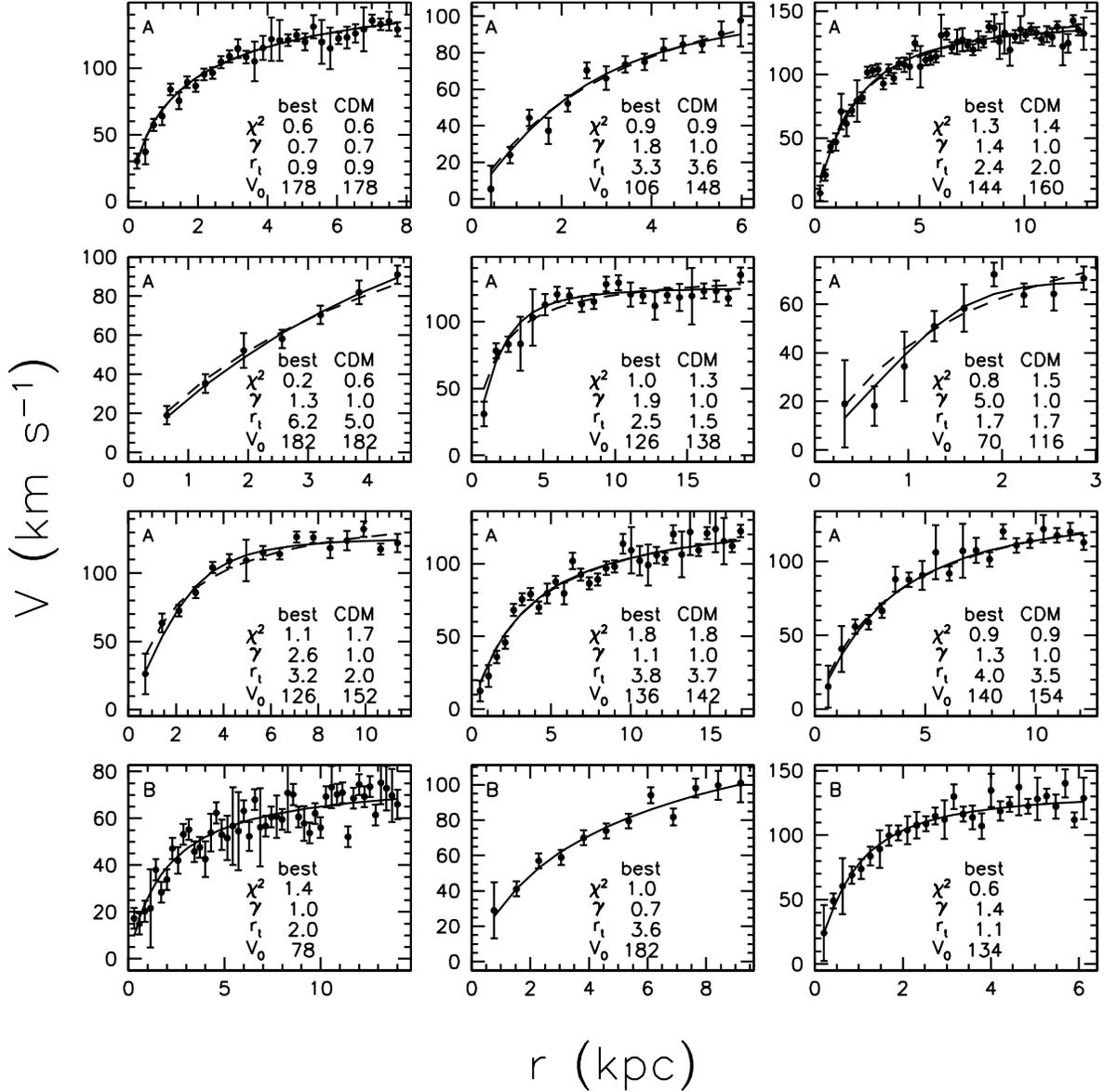}
\caption{Artificial ISO halo rotation curves with observational
  uncertainties added.  The top three rows show haloes for which a
  best fit could be found with Eq.~(1) as well as a constrained
  CDM-compatible fit. The fit parameters are listed in each panel. The
  full curve shows the best fit, the dashed curve the constrained CDM
  fit.  The bottom row shows three examples of curves for which a best
  fit could be found, but no CDM-compatible fit. In almost every case
  this was due to the ISO halo not meeting density criterion (iv).
\label{fig:panel}} \end{figure}

\begin{figure}
\plotone{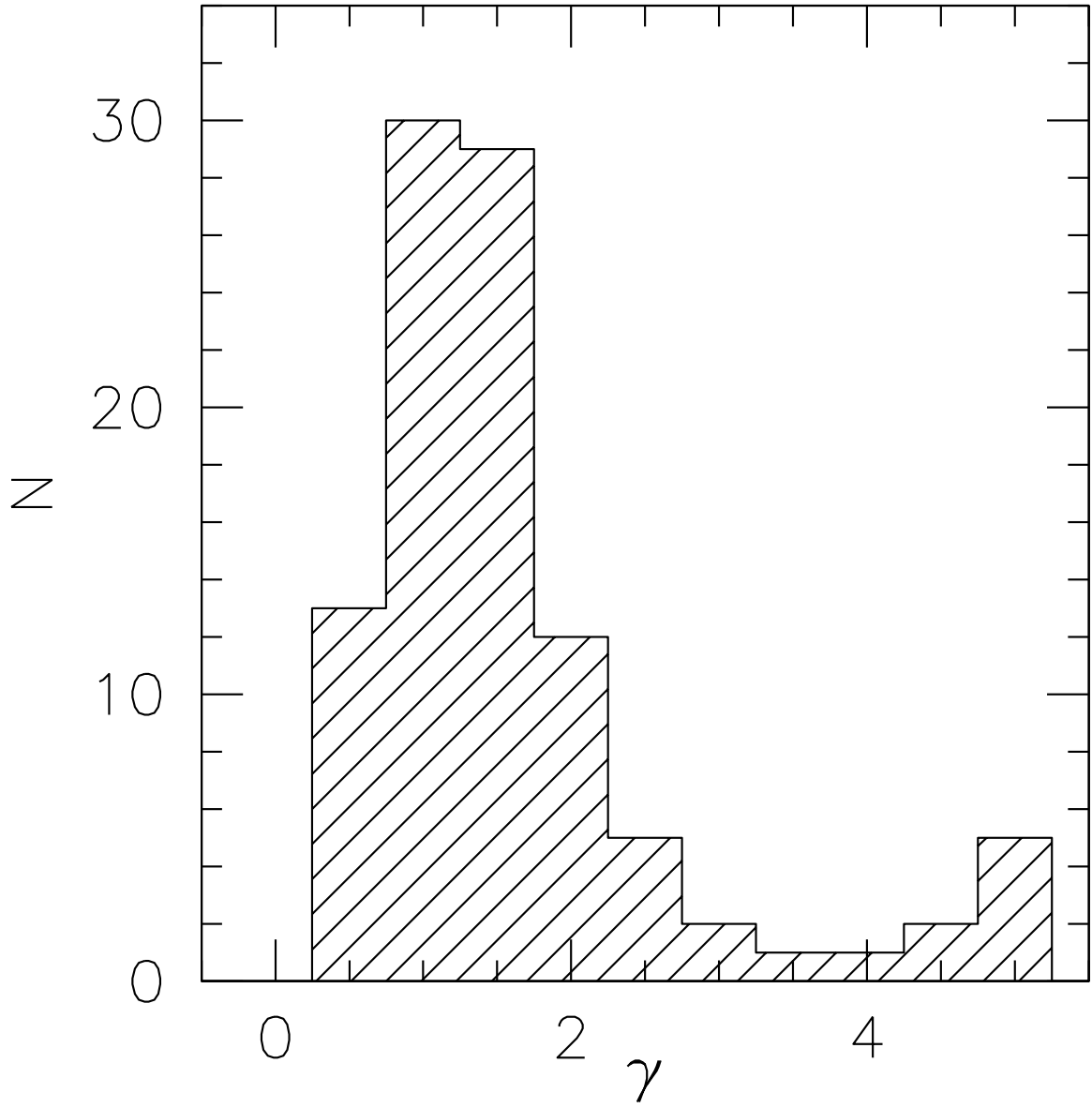}
\caption{Histogram of the best-fit $\gamma$ parameters of the ISO models.
\label{fig:modelgamma}} \end{figure}

\begin{figure}
\plotone{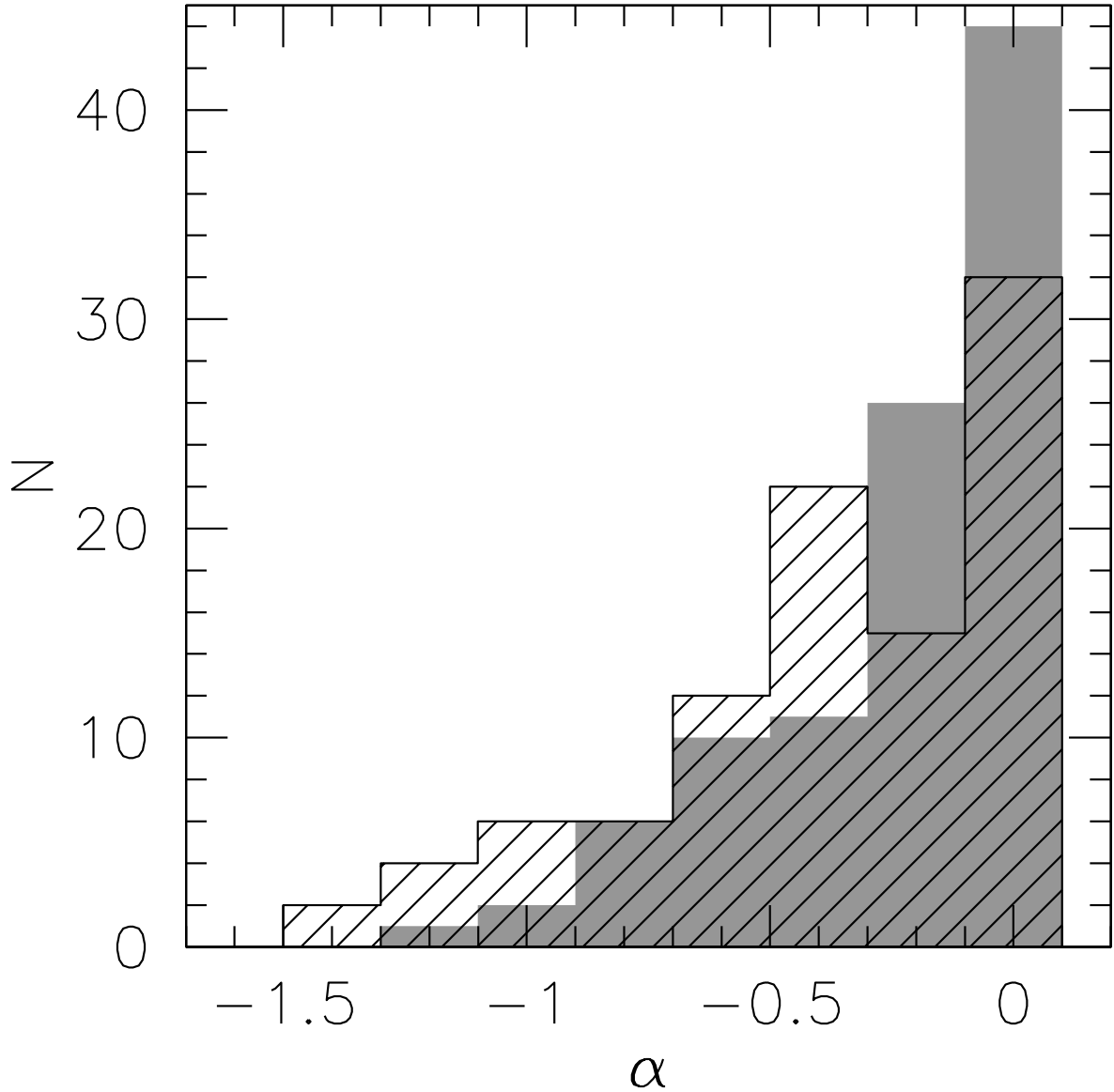} \caption{Histograms of the inner slopes at $r=0.4$ kpc of
  the input ISO halos (grey filled histogram), and the ``observed'' output slope at the same radius derived from the best $(\gamma,r_t)$-fit 
  (open hatched histogram). The larger range in slopes is larger than in Fig.~\ref{fig:hayslope3} due to the larger
range in core-radii used.
\label{fig:modelslopes}} \end{figure}

\begin{figure}
\plotone{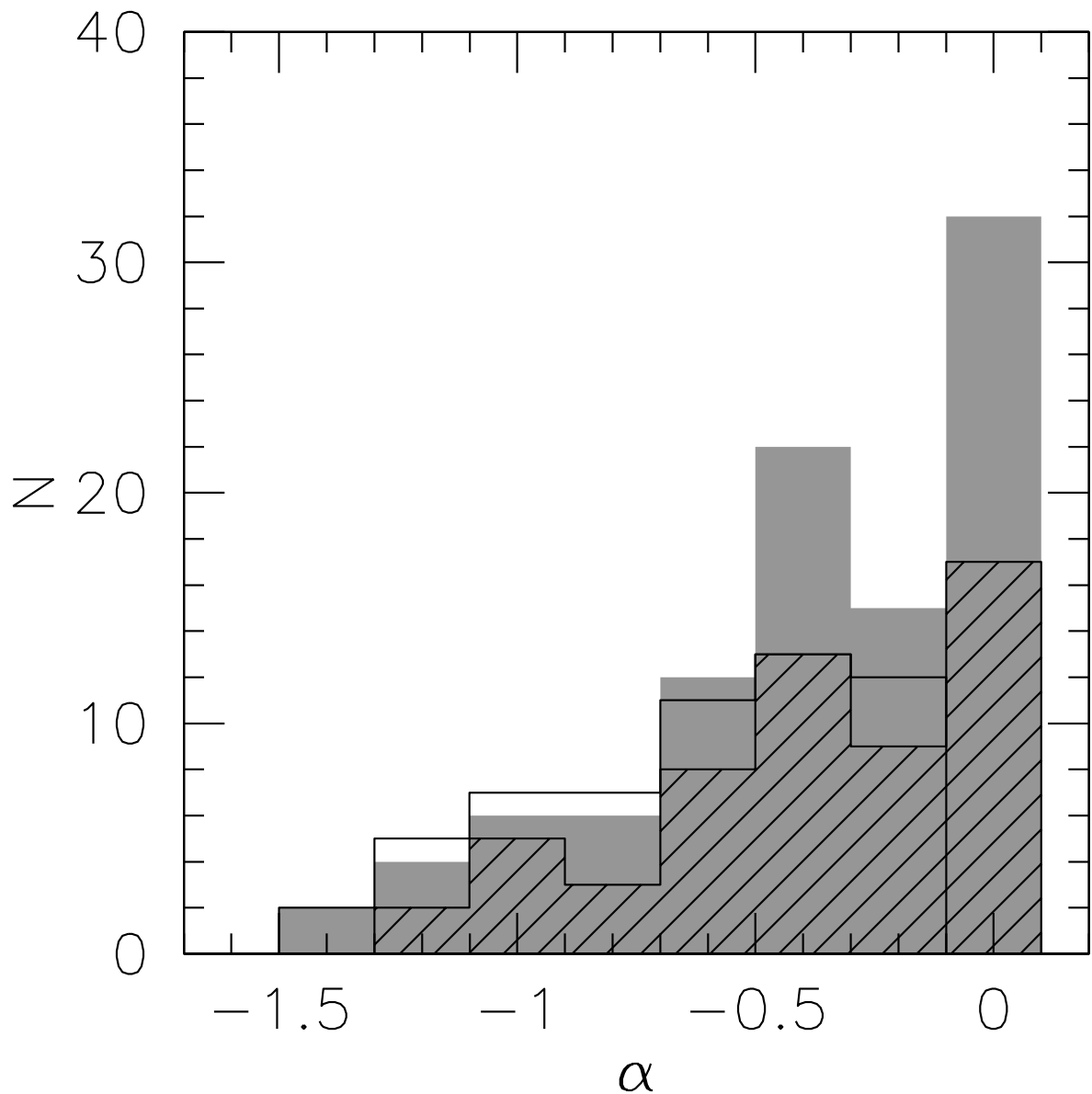} \caption{Histograms of the observed slopes at
  $r=0.4$ kpc of the model ISO haloes. The grey histogram shows the
  best-fit slope of the entire sample, the hatched histogram the
  best-fit slopes of the CDM-compatible haloes, and the open histogram
  shows the slopes of the CDM-compatible haloes derived using the
  CDM-compatible $\gamma$ and $r_t$ values.
\label{fig:slopescomp3}} \end{figure}

\begin{figure}
\plotone{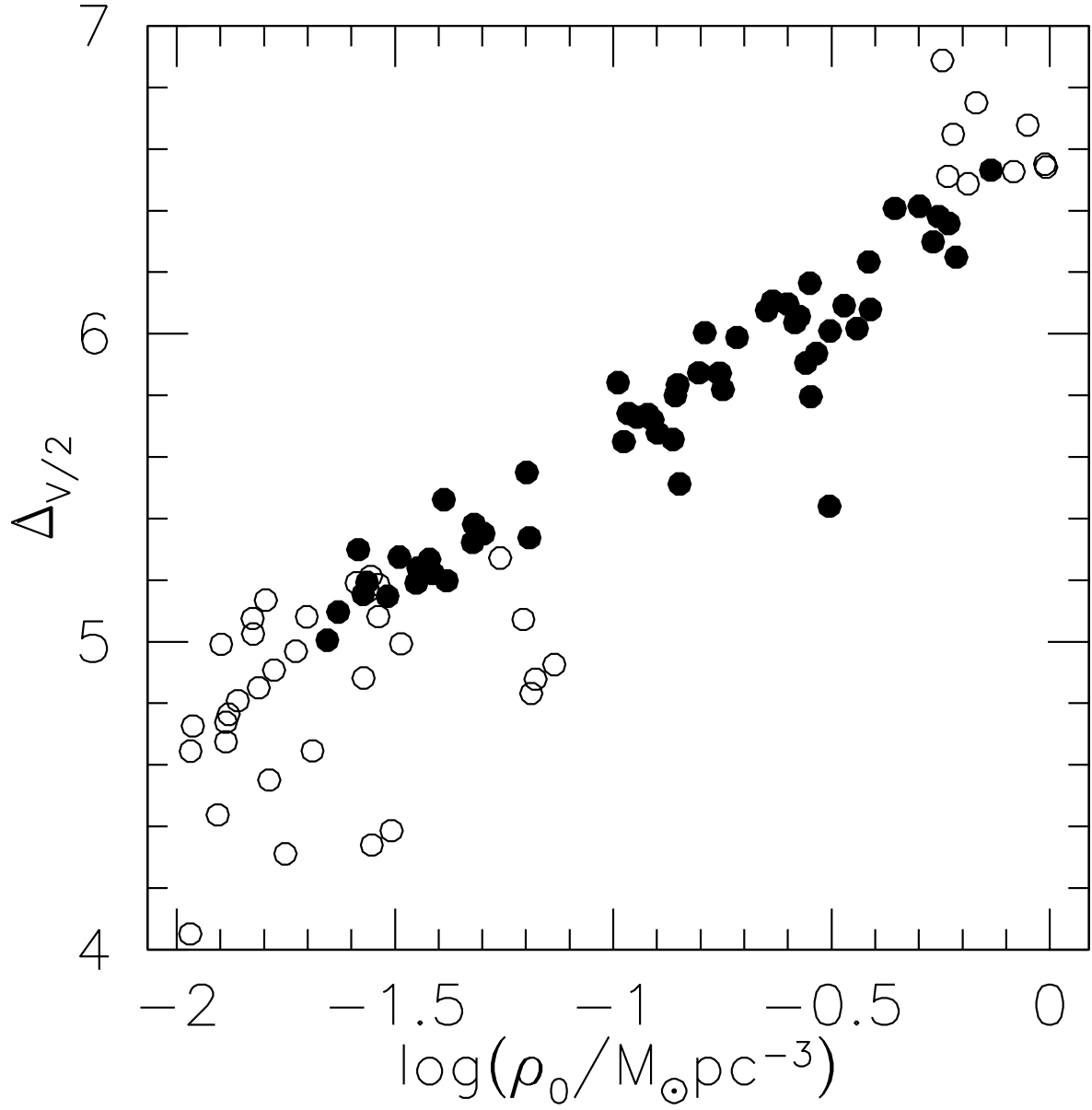} \caption{Average density $\Delta_{V/2}$ of a
  sample of ISO haloes plotted against the input central ISO density
  $\rho_0$. Filled points obey density criterion (iv), open points do
  not.
\label{fig:rho0}} \end{figure}

\begin{figure}
\plotone{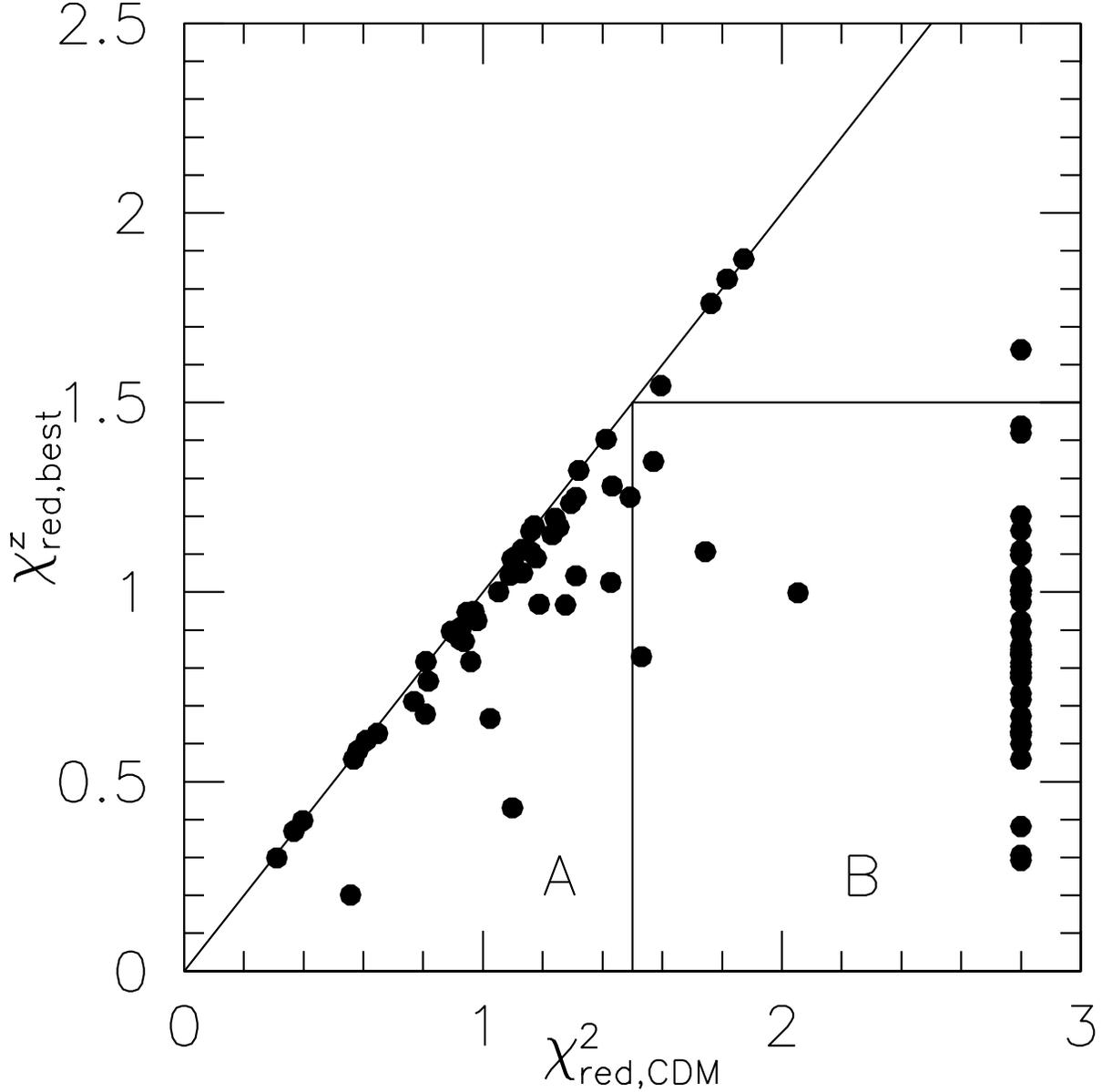} \caption{Comparison of the $\chi^2$ values of the
  best fits to the ISO haloes ($\chi^2_{\rm best}$, vertical axis),
  with those of the constrained CDM-compatible fits ($\chi^2_{\rm
    CDM}$, horizontal axis). Area A with $\chi^2_{\rm best} < 1.5$ and
  $\chi^2_{\rm CDM} < 1.5$ contains CDM-compatible ISO haloes. Area B
  with $\chi^2_{\rm best} > 1.5$ and $\chi^2_{\rm CDM} < 1.5$ contains
  ISO haloes for which good CDM fits could not be found. Haloes for
  which no CDM fit is defined due to criterion (iv) (see text) have
  for plotting purposes been given an arbitrary value of $\chi^2_{\rm
    CDM} = 2.8$.
\label{fig:chi2}} \end{figure}

\begin{figure}
\plotone{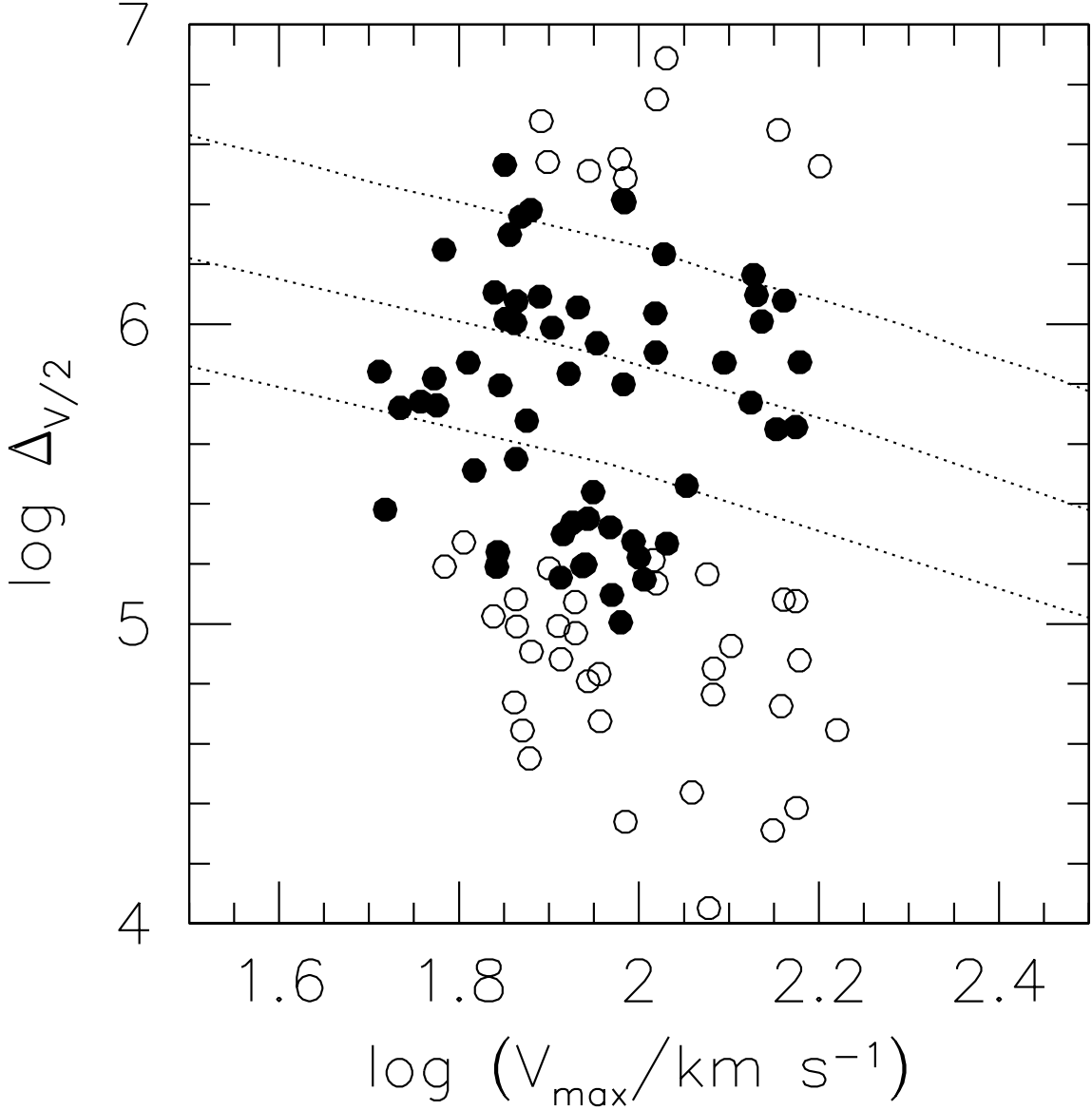}
\caption{Values of $\Delta_{V/2}$, as derived from the best $(\gamma,
  r_t, V_0)$ fit using Eq.~(1), plotted against the maximum rotation
  velocity $V_{\rm max}$ $(=V_{0,\rm best})$ for a sample of
  pseudo-isothermal (ISO) halo models. The filled points obey CDM
  criteria (i)-(iv) (see Sect.~1) and are thus supposed to be
  ``consistent'' with CDM in terms of shape and density. The open
  points meet criteria (i)-(iii) and are thus ```consistent'' in terms
  of shape of the rotation curve only.
\label{fig:bullock}} \end{figure}

In order to identify galaxies that are really compatible with CDM, an
additional constraint is needed, in such a way that only
$(\gamma,r_t)$ combinations are allowed that yield a steep slope [as
  well as pass conditions (i)-(iv)].  In practice this means that
$(\gamma, r_t)$ parameter space needs to be searched for the minimum
$\chi^2$ value that still results in a steep slope\footnote{Obviously,
  as Eq.~(1) also depends on $V_0$, one needs to take that parameter
  into account as well. This is done by stepping through values of
  $V_0$ at fixed $\gamma$ and $r_t$, and choosing the value that
  yields the lowest $\chi^2$. Therefore $\chi^2(\gamma,r_t) =
  \min[\chi^2(\gamma,r_t,V_0)]$.}.  In order to compare directly with
the simulations the slopes are again evaluated at $r=0.4$ kpc.  At
this radius the H04 simulations show slopes $-1.3 \la \alpha \la -1$,
and strictly speaking one ought to restrict the search of $(\gamma,
r_t)$ space to slopes $\alpha < -1$. However, to take into account
uncertainties in data and simulations, and give the CDM models as much
leeway as possible, a more liberal range of $\alpha < -0.8$ will be
used. Fig. \ref{fig:slopeonly} shows that this is the steepest slope
we can expect at this radius from an ultra-compact ISO halo with $R_C
\sim 0.5$ kpc, and this value thus cleanly separates the ISO and CDM
domains.

At this point a short digression on the significance of $\chi^2_{\rm
  red}$ is useful. In testing the goodness-of-fit of a model, the
reduced $\chi^2$, that is, the total $\chi^2$ divided by the degrees
of freedom of the data, is often used. A value of $\chi^2_{\rm
  red}\sim 1$ is then taken as an indication that the model is a good
description of the data. The reduced $\chi^2$ is an appropriate
statistic when comparing different models to a single set of data, or
data with a similar number of degrees of freedom. An example would be
the testing of an NFW model versus an ISO model for a single rotation
curve, or a set of curves with a similar number of data points.

H04 determine for each LSB rotation curve the reduced $\chi^2$ for the
best-fitting $(\gamma,r_t)$ model, as well as the minimum $\chi^2_{\rm
  red}$ value using their CDM constraints.  They argue that for most
of the galaxies the CDM constraints only cause a small increase in the
\emph{reduced} $\chi^2$, and that therefore the CDM-constrained models
are not much worse than the best-fit models. This is not necessarily
so. Remember that the reduced $\chi^2$ is the $\chi^2$ divided by the
degrees of freedom of the data (the number of data points minus the
number of parameters in the model). It is easy to see that in order to
cause the same change in \emph{reduced} $\chi^2$ the change in
\emph{absolute} $\chi^2$ must be much larger for a rotation curve with
a large number of data points than for a rotation curve with a small
number of data points.  In other words, a small change in
$\chi^2_{\rm red}$ can be very significant (that is, unlikely to be
consistent with the uncertainties in the data points) for a curve with
a large number of data points, while it can be insignificant (entirely
consistent with the uncertainties) for a curve with only a small
number of data points.

The observed sample we are dealing with here contains a large number
of rotation curves, with a wide range in the number of data points,
varying from $\sim 10$ to over 350.  The relevant parameter when
judging whether a change in fit parameters yields a model that is
still consistent with the data is thus \emph{not} the reduced
$\chi^2$, but the \emph{change in absolute $\chi^2$}.  The confidence
level of any parameter combination is given by the corresponding
change in $\chi^2$ with respect to the minimum value.  For a model
with three free parameters $(\gamma,r_t,V_0)$ as used here, the
$(1,2,3)\sigma$ confidence intervals for any fit are given by $\Delta
\chi^2 = (3.5, 8.0, 13.9)$, respectively.

Using these criteria, a rotation curve can be defined to be consistent
with CDM if the best-fit values of $\gamma$ and $r_t$ lie within
$2\sigma$ of the closest $(\gamma,r_t)$ combination with a $\alpha <
-0.8$ slope.  In order for the fit to give any meaningful constraints,
one should furthermore require that the area enclosed by the $2\sigma$
confidence level is a small fraction of the total area of parameter
space investigated\footnote{The choice for $2\sigma$ is not
  crucial. We have tested the analysis with a $3\sigma$ level as
  well. Though the number of badly constrained fits increases, the
  relative proportions of CDM consistent \emph{vs} inconsistent fits
  does not change appreciably.}. If the latter is not the case, then any
conclusions regarding agreement or disagreement with CDM are only very
weak at best.

Here the Group A rotation curves, defined in H04 to be consistent with
their definition of CDM, are investigated again using the new
constraints\footnote{Here only the \citet{dBB02} and
  \citet{mcgaugh2001} data are investigated.  However, there is no
  reason to believe that the \citet{swaters2003} data set differs from
  the other two in any way.}. Groups B and C are already inconsistent
with CDM using the more relaxed H04 constraints, and will therefore
not be investigated further.  For each rotation curve from Group A the
minimum $\chi^2$ was found using Eq.~(1), and the change in $\chi^2$
necessary to be consistent with a slope $\alpha < -0.8$ was
determined.  The $\sigma$-confidence interval corresponding to this
$\Delta \chi^2$ was also calculated, as well as the ratio $R$ of the
area enclosed by the $2\sigma$ contour and the total area of $(\gamma,
r_t)$ parameter space searched $(\gamma: 0 \rightarrow 5;\ r_t: 0
\rightarrow 5$ kpc).  As described above, the ratio $R$ is used to
determine whether a solution is well-constrained or not. Small values
of $R$ indicate that the $2\sigma$ area is only a small fraction of
total parameter space, and therefore yield significant
solutions. Large values of $R$ indicate that the fit does not
constrain models in a significant way.

\begin{figure}
\epsscale{0.45}
\plotone{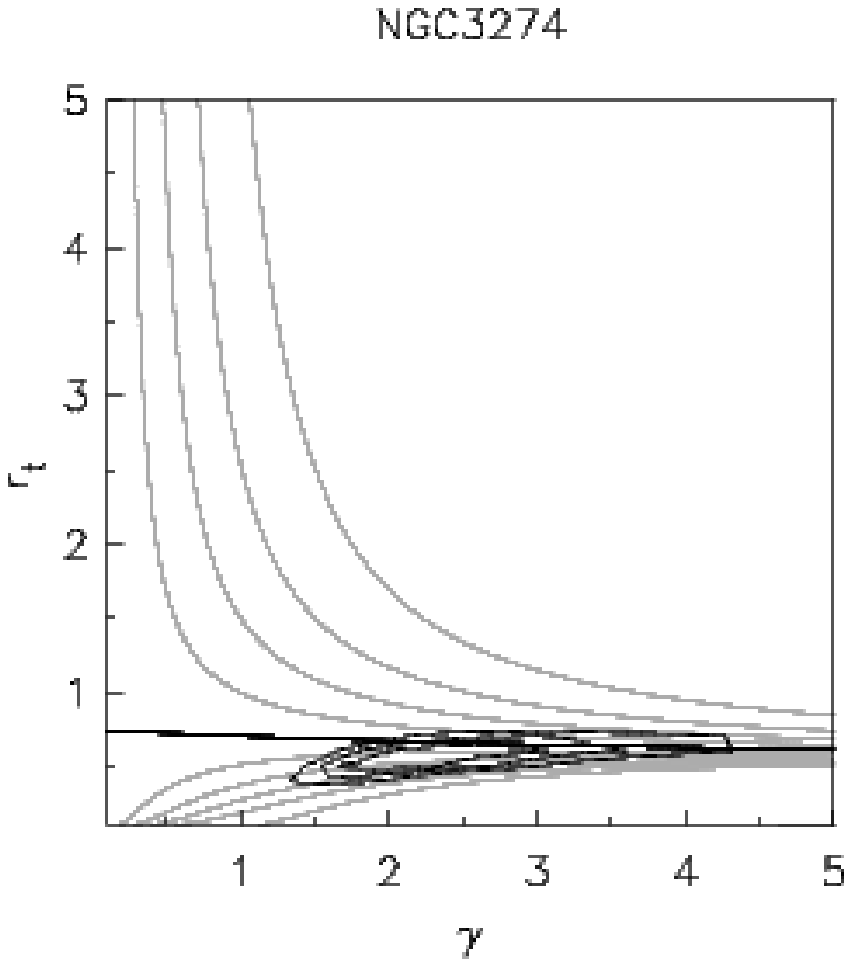}
\plotone{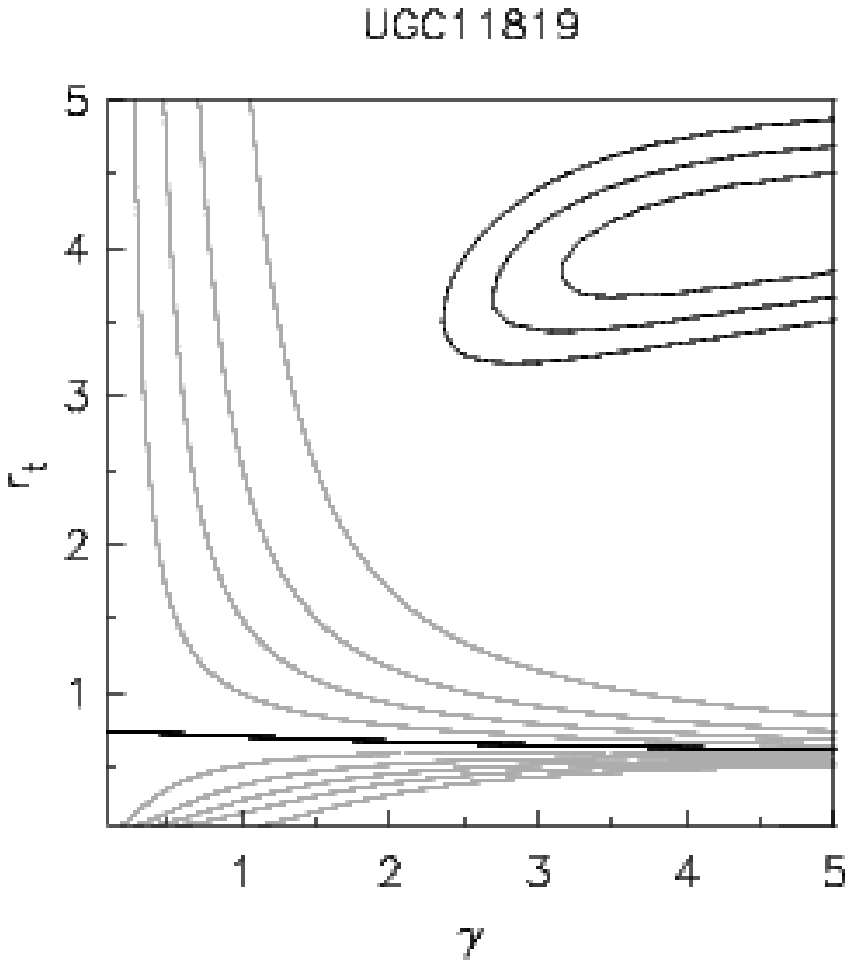}\\
\plotone{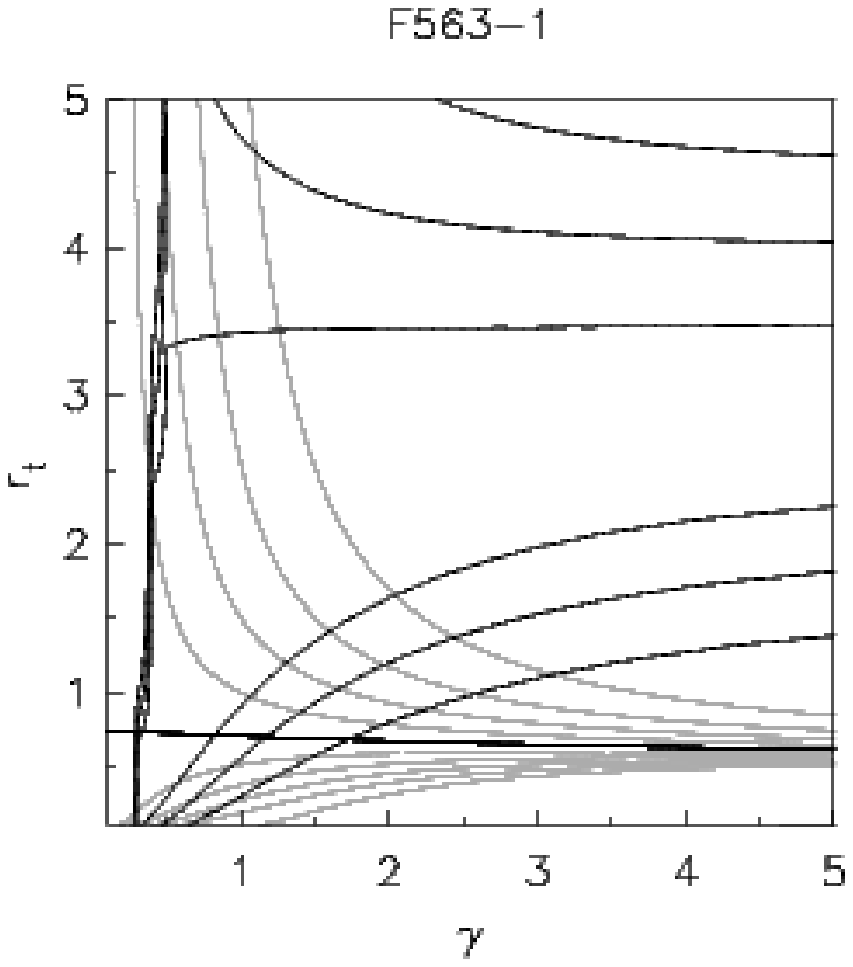}
\plotone{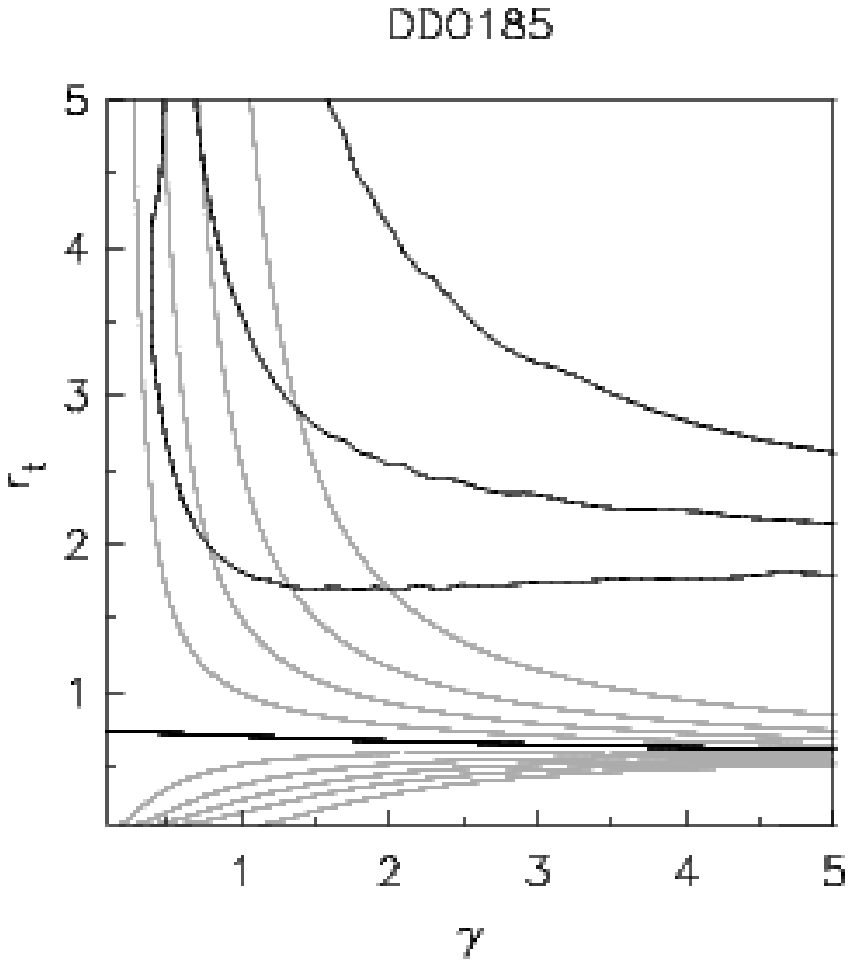}
 \caption{Contours of constant $\chi^2$ for representative LSB
   rotation curves. In all panels the contours indicate the
   $(1,2,3)\sigma$ confidence intervals.  \emph{Top-left:} NGC 3274,
   consistent within $2\sigma$, well-constrained $R=0.01$.
   \emph{Top-right:} UGC 11810, inconsistent at $>2\sigma$,
   well-constrained $R=0.07$. \emph{Bottom-left:} F563-1, consistent
   within $2\sigma$, badly constrained $R=0.56$. \emph{Bottom-right:}
   DDO 185, inconsistent at $>2\sigma$, badly-constrained $R = 0.44$.
\label{fig:chi2surface}} \end{figure}

\begin{figure}
\epsscale{1}
\plottwo{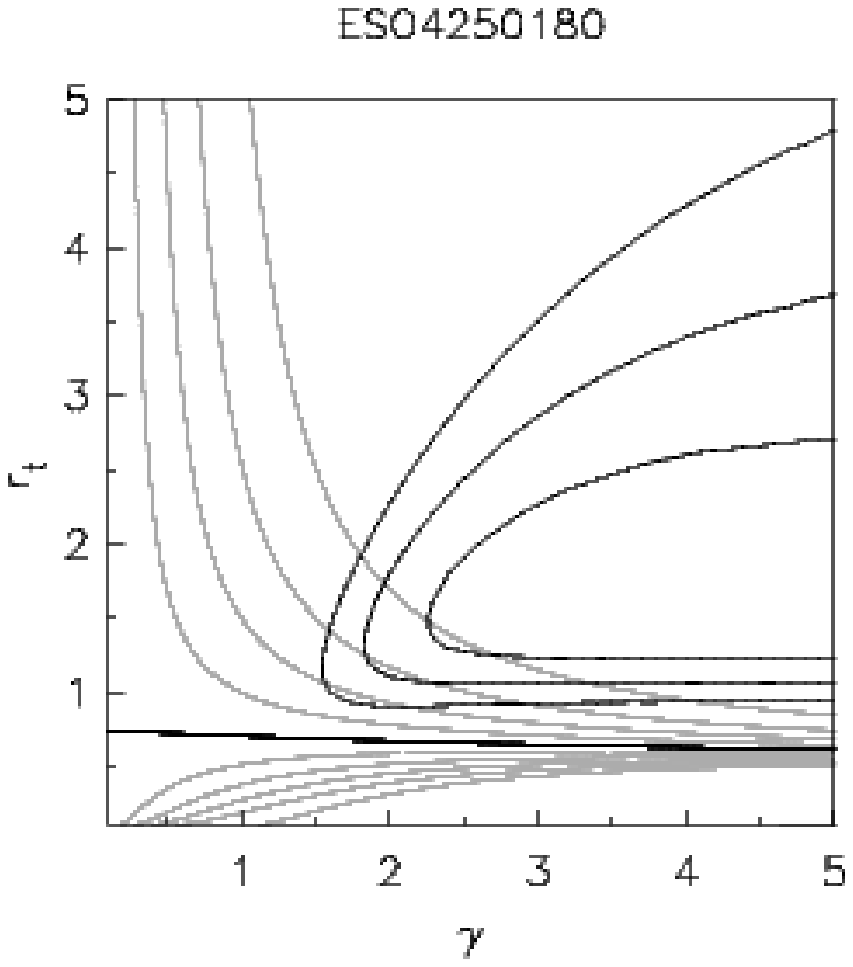}{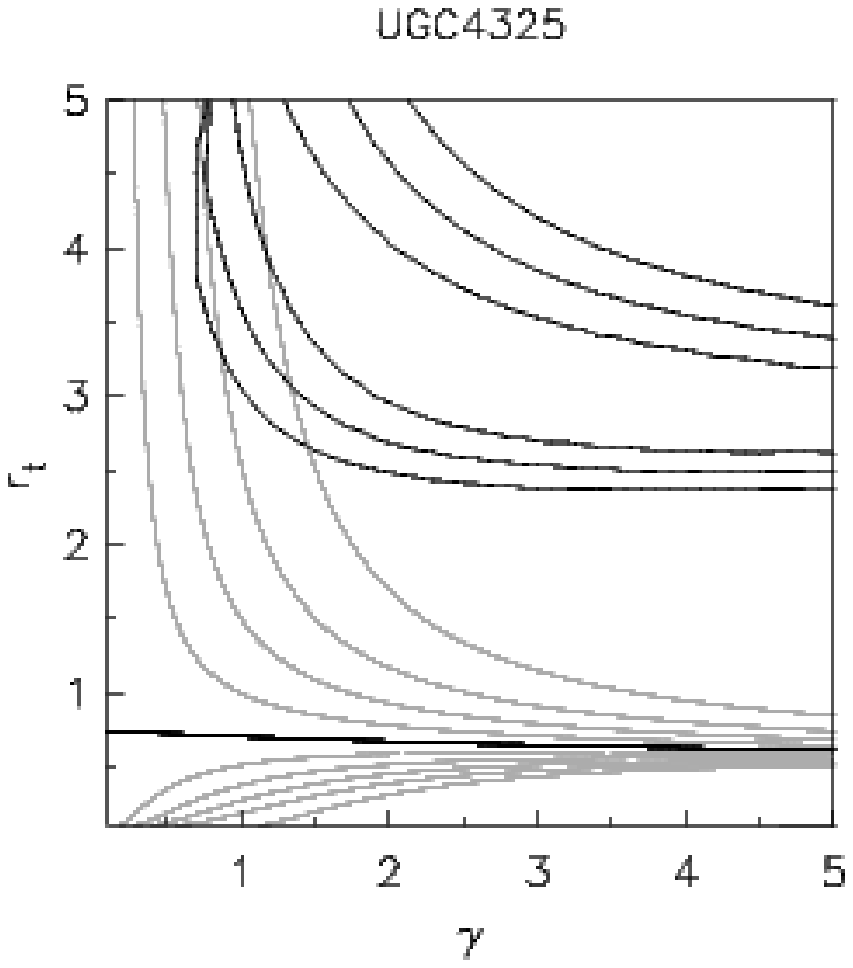}
 \caption{Two examples of galaxies with $R\simeq 0.25$.   In all panels the
contours indicate the $(1,2,3)\sigma$ confidence intervals.
\emph{Left:} UGC 4325. \emph{Right:} ESO 4250180.
\label{fig:chi2lim}} \end{figure}

\begin{figure}
\epsscale{1}
\plotone{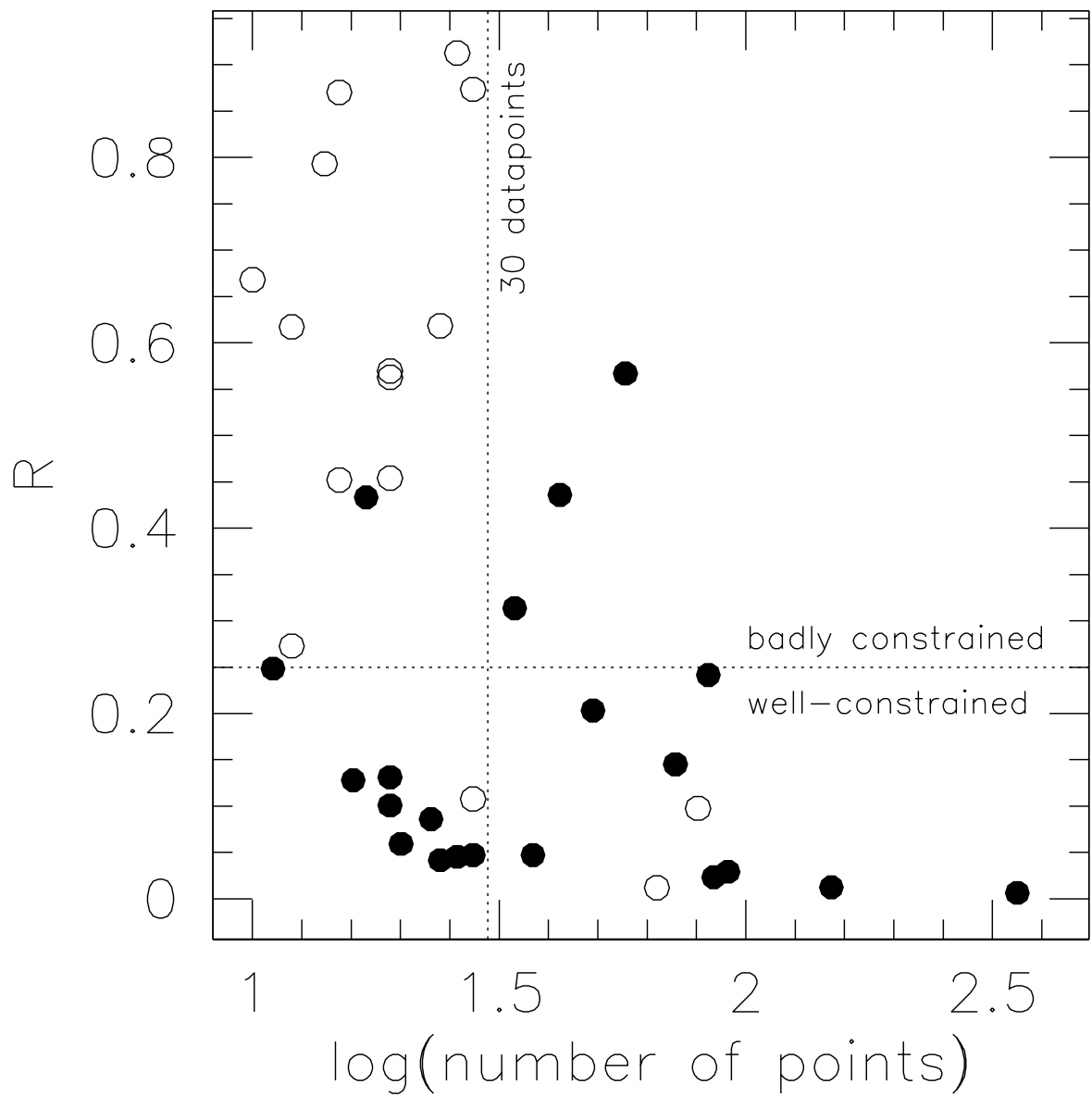}
\caption{The ratio $R$ of the area enclosed by the $2\sigma$ contour
  and the total parameter search area (see text for fuller
  description) plotted against the number of data-points in the
  rotation curve. Open circles have a best-fit $(\gamma,r_t)$ within
  $2 \sigma$ of a point with a steep slope $\alpha < -0.8$. Filled
  circles are more than $2\sigma$ away from such a point.
\label{fig:resolve}} 
\end{figure}

The results are given in Table \ref{table:fit}, where the galaxies are
sorted in order of the number of data-points in each individual
rotation curve.  Fig.~\ref{fig:chi2surface} shows the $(1,2,3)\sigma$
confidence contours for a number of representative galaxies as
described below.  From studying the plots of the $\chi^2$ contours for
all galaxies, it was found that a value $R=0.25$ is a reasonable
discriminator between well-constrained and badly constrained fits, but
the precise value of $R$ is not critical (see Fig.~\ref{fig:chi2lim}
for some examples of $R\simeq 0.25$ fits).  The galaxies in Table
\ref{table:fit} fall in four distinct categories:
\begin{enumerate}
\item \emph{Well-constrained solutions, consistent with CDM:} objects
  with best fits within $2\sigma$ of a steep slope $\alpha < -0.8$
  parameter combination, and with a well-constrained $R\le 0.25$
  solution.  An example is NGC3274, as shown in
  Fig.~\ref{fig:chi2surface}. This category consists of 3 out of the
  36 galaxies in Group A.
\item \emph{Well-constrained solutions, inconsistent with CDM:}
  objects with best fits more than $2\sigma$ away from a steep slope
  $\alpha < -0.8$ parameter combination, and with a well-constrained
  $R\le 0.25$ solution.  An example is UGC 11819, as shown in
  Fig.~\ref{fig:chi2surface}. This category contains 17 out of 36 galaxies.
\item \emph{Badly constrained solutions, consistent with CDM:} objects
  with best fits within $2\sigma$ of a steep slope $\alpha < -0.8$
  parameter combination, but  with a badly constrained $R > 0.25$
  solution.  An example is F563-1, as shown in Fig.~\ref{fig:chi2surface}. This category
  contains 12 out of 36 galaxies.
\item \emph{Badly constrained solutions, inconsistent with CDM:}
  objects with best fits more than $2\sigma$ away from a steep slope
  $\alpha < -0.8$ parameter combination, and with a badly constrained
  $R > 0.25$ solution.  An example is DDO 185, as shown in
  Fig.~\ref{fig:chi2surface}. This category contains 4 out of 36 galaxies.
\end{enumerate}
It is clear that only galaxies from the first two categories,
i.e.\ with $R\le 0.25$, can help constrain halo models.  A glance at
Table \ref{table:fit} shows that most of the solutions that are
consistent with CDM are actually badly constrained. This is
investigated further in Fig.~\ref{fig:resolve}.  Here $R$ is plotted
against the number of data points in each rotation curve. A
distinction is made between rotation curves that are consistent with
CDM and those that are inconsistent with CDM using the constraints
introduced here.  Table \ref{table:fit} and Fig.~\ref{fig:resolve}
lead to several conclusions. Firstly, of the 15 galaxies that are
consistent with CDM, 13 have less than 30 data points. Furthermore, of
these 13 galaxies, only 1 has $R\le 0.25$ (i.e.\ is well-constrained).
Looking at the 21 galaxies that are inconsistent with CDM, only 10
have less than 30 data points. Of these 10 galaxies, all except one
have $R\le 0.25$.  Rotation curves with more than 30 data points tend
to be better constrained on average, as well as have a tendency to be
inconsistent with CDM.  It is also interesting to note that even
amongst galaxies with only a limited number of data points, the most
constrained solutions (small $R$) tend to be those inconsistent with
CDM.  The majority of solutions $(\sim 85$ per cent) that are
consistent with CDM have only a small number of data points \emph{and}
are ill-constrained.  These ill-constrained solutions therefore do not
decide between any model one way or the other, and statements that
these galaxies are consistent with CDM are thus not very strong.  Of
the well-constrained solutions the large majority is inconsistent with
CDM.  Furthermore, of the 13 galaxies with more than 30 data points
only 2 are consistent with CDM.  These two galaxies (NGC 3274 and NGC
4455) also happen to have the smallest optical scale-lengths and
highest surface brightnesses from all galaxies with available
photometry in the \citet{mcgaugh2001}, \citet{dBB02} and
\citet{swaters2003} samples: both galaxies have $\mu_{0,B} \la 20.0$
mag arcsec$^{-2}$.  These galaxies are therefore unlikely to be
dominated by dark matter in their inner parts and it is very likely
that the steep slopes in these two galaxies simply reflect the stellar
mass distribution (see also the discussion on NGC 3274 in Sec.~9.2.2
of \citealt{dBB02}).

Amongst the rotation curves with less than 30 points there is only one
galaxy with $R<0.25$ that seems consistent with CDM. This is UGC 731,
a bona-fide LSB dwarf, and is the best and only candidate in the
entire observed sample for a CDM-consistent, dark-matter-dominated
galaxy with a well-constrained solution. The level of agreement is not
perfect though, UGC 731 is consistent with a $\alpha = -0.8$ slope at
the $1.3\sigma$ confidence level, with the confidence level decreasing
to $1.7\sigma$ for an $\alpha = -1$ slope, which is still only the
upper limit of the range of slopes expected from the simulations.

Once the steep slope constraint is added to the H04 constraints there
are thus only 3 galaxies left that have a well-constrained fit
consistent with a steep slope and $\Lambda$CDM. Of these 3 galaxies, 2
are of very high surface brightness and are likely to be dominated by
stars in the inner parts, leaving only one possible candidate with a
well-constrained CDM-compatible fit.  The large majority of the
well-constrained fits strongly prefer shallow slopes.

The 2$\sigma$ limit used here already indicates that for many galaxies
it is difficult to adjust the fit parameters to get a CDM-compatible
fit.  Restrictive as this criterion is, it really only gives an upper
limit to the true probability. While one has the freedom to adjust the
parameters of an individual fit to test the likelihood of an
hypothesis for one particular rotation curve, this becomes problematic
when dealing with many fits. If every fit is individually adjusted to
fit a particular model, does that still constitute a proper test? An
analogue can perhaps be found in some of the first supernova results
that indicated the existence of a cosmological constant (e.g.\
\citealt{riess}). The majority of the supernovae in that early work
were each individually consistent within $2\sigma$ with the
\emph{non-}existence of $\Lambda$. Even though the data showed a
systematic trend, each measurement could have been corrected
individually to fit the no-$\Lambda$ model. However, the joint result
of these small inconsistencies has formed the basis for the discovery
of the effects of $\Lambda$ on the expansion of the universe. While it
would be presumptuous to compare those results with the current
analysis, it does show that it is important to 
pay attention to the global conclusion that is forced upon us by the
data.

One thing that is clear is that galaxies that are significantly
consistent with $\Lambda$CDM in terms of their inferred central
densities and the shapes of their rotation curves seem to be rare
indeed.

\section{Summary}

I have investigated the claim by H04 that the majority of rotation
curves of LSB galaxies are not inconsistent with CDM.  I have shown
that the method used is unable to distinguish between shallow and
steep slopes, even when these are obviously present and easily
detectable. I illustrate this by showing that with the H04 method one
would infer even pseudo-isothermal, core-dominated haloes to be
consistent with CDM.  An extra constraint based on the mass-density
slope is thus needed to make more definitive statements on any
(dis)agreement with CDM.  Using this extra criterion I show that only
a quarter of the LSB galaxies can be said to be consistent with CDM,
as opposed to the three quarters found by H04. The majority of these
CDM-consistent LSB galaxies do however have fit parameters that are so
ill-constrained, that they are consistent with virtually
anything. Restricting the analysis to galaxies with well-constrained
solutions, I find that only 3 out of the 20 well-constrained galaxies
are possibly consistent with CDM; two of these are High Surface
Brightness dwarf galaxies that are likely to be dominated by stars.

In summary, a comparison of the circular velocity profiles of CDM
haloes with rotation curves of LSB galaxies indicates that the shapes
and inferred central densities of most LSB galaxies are mostly
\emph{inconsistent} with those of simulated haloes within the
limitations imposed by observational error.

\acknowledgements It is a pleasure to thank Charles Jenkins, Stacy McGaugh, Chris Power
and David Weldrake for useful discussions.  I thank the anonymous
referee for constructive comments.

\begin{deluxetable}{lrrrrl}
\tablewidth{0pt}
\tablecaption{Fit parameters LSB rotation curves\label{table:fit}}

\tablehead{\colhead{Name} & \colhead{Points} & \colhead{$\Delta \chi^2$} & \colhead{$R$} & \colhead{$\sigma$} & \colhead{Ref.}\\ 
\colhead{(1)} & \colhead{(2)} & \colhead{(3)} & \colhead{(4)} & \colhead{(5)} & \colhead{(6)} 
} 
\startdata
F563-1 & 10 & 1.92 & 0.67 & {0.54} & M01\\
ESO 4250180 & 11 & 24.94 & 0.25 & 4.31 & M01 \\
ESO 4880490 & 12 & 1.07 & 0.27 & {0.28} & M01\\
ESO 3020120 & 12 & 1.72 & 0.62 & {0.48} & M01\\
DDO 189 & 14 & 0.75 & 0.79 & {0.18} &dB02\\
ESO 1870510 & 15 & 0.02 & 0.45 & 0.0011 &M01\\
UGC 4115 & 15 & 0.36 & 0.87 & 0.064 &M01\\
F568-3 & 16 & 43.27 & 0.13 & 5.99 &M01\\
F583-1 & 17 & 15.12 & 0.43 & 3.14 &M01\\
F571-8 & 19 & 93.60 & 0.10 & $>7$ & M01 \\
F563-1 & 19 & 0.02 & 0.56 & 0.0011 & dB02 \\
UGC 11583 & 19 & 4.31 & 0.57 & 1.20 & M01 \\
ESO 1200211 & 19 & 2.09 & 0.45 & 0.59 & M01 \\
F583-4 & 19 & 19.67 & 0.13 & 3.72 & M01 \\
ESO 2060140 & 20 & 9.57 & 0.06 & 2.28 & M01 \\
UGC 11819 & 23 & 181.61 & 0.07 & $>7$ & M01 \\
F730-V1 & 24 & 18.66 & 0.04 & 3.60 & M01 \\
UGC 4173 & 24 & 0.00 & 0.62 & 0.00 & dB02 \\
UGC 11616 & 26 & 105.32 & 0.05 & $>7$ & M01 \\
DDO 47 & 26 & 0.05 & 0.91 & 0.0040 & dB02 \\
UGC 731 & 28 & 4.95 & 0.11 & 1.35 & dB02 \\
UGC 11454 & 28 & 218.96 & 0.05 & $>7$ & M01 \\
UGC 5005 & 28 & 1.47 & 0.87 & 0.40 & dB02 \\
UGC 1230 & 34 & 20.27 & 0.31 & 3.79 & dB02 \\
UGC 628 & 37 & 8.92 & 0.05 & 2.17 & dB02 \\
DDO 185 & 42 & 19.23 & 0.44 & 3.67 & dB02 \\
UGC 10310 & 49 & 32.78 & 0.20 & 5.09 & dB02 \\
UGC 3371 & 57 & 11.68 & 0.57 & 2.63 & dB02 \\
NGC 3274 & 66 & 0.00 & 0.01 & 0.00 & dB02 \\
UGC 3137 & 72 & 24.17 & 0.15 & 4.23 & dB02 \\
NGC 4455 & 80 & 1.63 & 0.10 & 0.45 & dB02 \\
UGC 4325 & 84 & 76.41 & 0.24 & $>7$ & dB02 \\
DDO 52 & 86 & 41.8 & 0.02 & 5.87 & dB02 \\
NGC 100 & 92 & 21.16 & 0.03 & 3.90 & dB02 \\
NGC 5023 & 149 & 95.96 & 0.01 & $>7$ & dB02 \\
NGC 4395 & 355 & 23.51 & 0.01 & 4.16 & dB02 \\
\enddata
\tablecomments{(1) Identification of galaxy. (2) Number of data-points in rotation curve. (3) $\Delta \chi^2$ needed to find solution with slope $\alpha < -0.8$. (4) Ratio of area enclosed by 2$\sigma$ contour and total search area. (5) Confidence level of $\Delta \chi^2$ needed to find solution with steep slope. (6) Source of the data: dB02: \citet{dBB02}; M01: \citet{mcgaugh2001}.}



\end{deluxetable}

\clearpage

\begin{thebibliography}{}
\bibitem[Alam et al.(2002)]{alam02} Alam, S.~M.~K., Bullock, 
J.~S.,   Weinberg, D.~H.\ 2002, \apj, 572, 34 
\bibitem[Binney(2004)]{binney2004} Binney, J.\ 2004, IAU 
Symposium, 220, 3 
\bibitem[Bullock et al.(2001)]{bullock01} Bullock, J.~S., Kolatt, 
T.~S., Sigad, Y., Somerville, R.~S., Kravtsov, A.~V., Klypin, A.~A., 
Primack, J.~R.,\&  Dekel, A.\ 2001, \mnras, 321, 559 
\bibitem[Courteau(1997)]{courteau} Courteau, S.\ 1997, \aj, 114, 
2402 
\bibitem[de Blok(2004)]{blok2004} de Blok, W.~J.~G.\ 2004, IAU 
Symposium, 220, 69 
\bibitem[de Blok \&  Bosma(2002)]{dBB02} de Blok, W.~J.~G.,  
Bosma, A.\ 2002, \aap, 385, 816 
\bibitem[de Blok \& McGaugh(1997)]{blok97} de Blok, W.~J.~G., 
McGaugh, S.~S.\ 1997, \mnras, 290, 533 
\bibitem[de Blok et al.(2001a)]{blok2001a} de Blok, W.J.G., 
McGaugh, S.S., Rubin, V.C. 2001, \aj, 122, 2396
\bibitem[de Blok et al.(2001b)]{blok2001b} de Blok, W.~J.~G., 
McGaugh, S.~S., Bosma, A., \& Rubin, V.~C.\ 2001, \apj, 552, L23
\bibitem[de Blok et al.(2003)]{blok2003} de Blok, W.~J.~G., 
Bosma, A., \& McGaugh, S.\ 2003, \mnras, 340, 657 
\bibitem[Fukushige et al.(2004)]{fuku} Fukushige, T., Kawai, 
A., \& Makino, J.\ 2004, \apj, 606, 625 
\bibitem[Ghigna et al.(2000)]{ghigna} Ghigna, S., Moore, B., 
Governato, F., Lake, G., Quinn, T., \& Stadel, J.\ 2000, \apj, 544, 616 
\bibitem[Hayashi et al.(2004)]{hayashi} Hayashi, E., Navarro, J.F., Power, 
C., Jenkins, A., Frenk, C.S., White, S.D.M., Springel, V., Stadel, J.,
Quinn, T.R. 2004, \mnras, 355, 794
\bibitem[Jing et al.(1995)]{jing} Jing, Y.~P., Mo, H.~J., 
Borner, G., \& Fang, L.~Z.\ 1995, \mnras, 276, 417 
\bibitem[Klypin et al.(2001)]{klypin} Klypin, A., Kravtsov, 
A.~V., Bullock, J.~S., \& Primack, J.~R.\ 2001, \apj, 554, 903
\bibitem[McGaugh et al.(2001)]{mcgaugh2001} McGaugh, S.~S., Rubin, 
V.~C., \& de Blok, W.~J.~G.\ 2001, \aj, 122, 2381
\bibitem[Moore et al.(1998)]{moore98} Moore, B., Governato, F., 
Quinn, T., Stadel, J., \& Lake, G.\ 1998, \apj, 499, L5 
\bibitem[Moore et al.(1999)]{moore99} Moore, B., Ghigna, S., 
Governato, F., Lake, G., Quinn, T., Stadel, J., \& Tozzi, P.\ 1999, \apj, 
524, L19 
\bibitem[Navarro, Frenk \& White(1996)]{NFW96} Navarro, J.F., Frenk, C.S.,  
White, S.D.M. 1996, \apj, 462, 563
\bibitem[Navarro, Frenk \& White(1997)]{NFW97} Navarro, J.F., Frenk, C.S., 
White, S.D.M. 1997, \apj, 490, 493
\bibitem[Navarro et al.(2004)]{NFW04} Navarro, J.~F., et al.\ 
2004, \mnras, 349, 1039 
\bibitem[Power et al.(2003)]{power03} Power, C., Navarro, 
J.~F., Jenkins, A., Frenk, C.~S., White, S.~D.~M., Springel, V., Stadel, 
J., \& Quinn, T.\ 2003, \mnras, 338, 14 
\bibitem[Riess et al.(1998)]{riess} Riess, A.~G., et al.\ 
1998, \aj, 116, 1009 
\bibitem[Swaters et al.(2003)]{swaters2003} Swaters, R.~A., Madore, 
B.~F., van den Bosch, F.~C., \& Balcells, M.\ 2003, \apj, 583, 732 
\bibitem[Wechsler et al.(2002)]{wechsler02} Wechsler, R.~H., 
Bullock, J.~S., Primack, J.~R., Kravtsov, A.~V., \& Dekel, A.\ 2002, \apj, 
568, 52 
\end{thebibliography}
\end{document}